\documentclass[11pt,epsf,letterpaper]{article}
\usepackage{setspace}
\usepackage{color}
\usepackage{amsmath}
\usepackage{amsfonts}
\usepackage{verbatim}
\usepackage{amssymb}
\usepackage{graphicx}
\usepackage{epstopdf}
\usepackage{epsf}
\usepackage{bbm}
\usepackage[caption=false]{subfig}
\usepackage{placeins}

\definecolor{darkgreen}{rgb}{0,0.35,0}

\providecommand{\U}[1]{\protect\rule{.1in}{.1in}}
\begin{document}

\title{Analytic Studies of Static and Transport Properties of (Gauged)
Skyrmions}
\author{F. Canfora$^{1}$, N. Dimakis$^{2}$, A. Paliathanasis$^{3,4}$ \\
$^{1}$\textit{Centro de Estudios Cient\'{\i}ficos (CECS), Casilla 1469,
Valdivia, Chile.}\\
$^{2}$\textit{Center for Theoretical Physics, College of Physical Science
and Technology,}\\
\textit{Sichuan University, Chengdu 610065, China.}\\
$^{3}$\textit{Instituto de Ciencias F\'isicas y Matem\'aticas, Universidad
Austral de Chile, Valdivia, Chile.}\\
$^{4}$\textit{Institute of Systems Science, Durban University of Technology,}%
\\
\textit{PO Box 1334, Durban 4000, Republic of South Africa.}\\
{\small canfora@cecs.cl, nsdimakis@scu.edu.cn, anpaliat@phys.uoa.gr}}
\maketitle

\begin{abstract}
We study static and transport properties of Skyrmions living within a finite
spatial volume in a flat (3+1)-dimensional spacetime. In particular, we
derive an explicit analytic expression for the compression modulus
corresponding to these Skyrmions living within a finite box and we show that
such expression can produce a reasonable value. The gauged version of these
solitons can be also considered. It is possible to analyze the order of
magnitude of the contributions to the electrons conductivity associated to
the interactions with this Baryonic environment. The typical order of
magnitude for these contributions\ to conductivity can be compared with the
experimental values of the conductivity of layers of Baryons.
\end{abstract}

\section{Introduction}

The appearance of Skyrme theory \cite{skyrme} disclosed very neatly the
fundamental role of topology in high energy physics (see for instance \cite%
{nuc0,nuc1,nuc2,nuc3,nuc4,nuc5}). First of all, the low energy QCD is very
well described by the Skyrme theory \cite{witten0}. Secondly, the solitons
of this Bosonic theory (\textit{Skyrmions}) describe Baryons. Thirdly, the
Baryon charge is the winding number of the configuration (see \cite%
{witten0,finkrub,manton,skyrev1,giulini,bala0,ANW,guada} and references
therein).

These arguments are more than enough to justify a profound analysis of the
Skyrme model. Indeed, extensive studies of the latter can be found in
literature (as the previous references clearly show). Not surprisingly%
\footnote{%
At least taking into account that it is reasonable to expect that the theory
describing the low energy limit of QCD should be a quite complicated one.},
the Skyrme field equations are a very hard nut to crack and, until very
recently no analytic solution was available. Nevertheless, many numerical
studies have shown that the Skyrme model provides results in good agreement
with experiments.

Despite the success of the model and the existence of several solutions among different contexts, the analysis of their  phenomenological aspects seldom can be carried out in an analytic manner. For an analytic solution and a relevant study in compact manifolds see \cite{newref1}.

The gauged Skyrme model (which describes the coupling of a $U(1)$ gauge
field with the Skyrme theory) has also very important applications in the
analysis of electromagnetic properties of Baryons, in the decay of nuclei in
presence of defects (see \cite%
{witten0,Witten,gipson,goldstone,dhoker,rubakov} and references therein).
Obviously, from the point of view of constructing analytic solutions, the $%
U(1)$ gauged Skyrme model is even worse than the original Skyrme theory.
Until very recently, no explicit topologically non-trivial solution was
available. Thus, topological configurations of this theory have been deeply
analyzed numerically (see \cite{gaugesky1,gaugesky2}\ and references
therein).

Here we list three relevant problems in the applications of (gauged) Skyrme
theory to high energy phenomenology which will be the focus of the present
paper.

\textbf{1)} \textit{Finite density effects and the compression modulus}:
Finite density effects (and, in general, the phase diagrams) in the
Skyrme model have been historically a very difficult topic to analyze with
analytic methods. The lack of explicit solutions with topological charge
living within a finite flat box with the spherical Skyrme ansatz is the
origin of the problem. Some numerical results with the use of the spherical
Skyrme ansatz are presented in \cite%
{klebanov,chemical1,chemical2,chemical3,chemical4} and references therein.
Due to the fact that both finite volume effects and isospin chemical
potential break spherical symmetry it is extremely difficult to improve the
pioneering results in \cite{klebanov,chemical1,chemical2,chemical3,chemical4}%
\ without changing the original Skyrme ansatz. The main problem in this
group is certainly the \textit{compression modulus} \cite%
{probcompression1,probcompression2,probcompression3} (to be defined
precisely in the next section) which, roughly speaking, has to do with the
derivative of the total energy of the Skyrmions with respect to the volume.
The experimental value is different from the value derived using the
original spherical hedgehog ansatz. The usual way to compute the compression
modulus is to assume the Derrick rescaling for the reaction of nuclear
matter to the action of external pressure (see the detailed discussion in
\cite{Adam}). The resulting value is higher than the experimental value%
\footnote{%
The following analysis suggests that this "uniform rescaling" assumption
could be too strong. Indeed, the results at the end of section 3 shows that
Skyrme theory, when analyzed at finite density, provides with values of the
compression modulus which are close to the experimental one.}. A closely
related technical difficulty is that, if one uses the original hedgehog
ansatz for the Skyrmion, it is very unclear even \textit{how to define} the
compression mod\qquad ulus since the original Skyrme ansatz describes a
spherical Skyrmion living within an infinite volume so that to compute the
derivatives of the energy with respect to the volume becomes a subtle
question. The best way out of this difficulty would be, of course, to have a
consistent ansatz for a Skyrmion living within a finite volume. Relevant numerical results in the literature on that problem are presented
in \cite{ref1,ref2,ref3,ref4} where non-spherical ans\"atze have been considered.

\textbf{2)} \textit{Existence of Skyrmion-antiSkyrmion bound
states/resonances}: multi-Skyrmionic bound states of Baryon charge higher
than 1 are known to exist and they have been successfully constructed
numerically (see, for instance, \cite{manton}\ and references therein).
However, until very recently, the problem of the existence of
Skyrmion-antiSkyrmion bound states and resonances did not possess the place
it deserved in the literature on the Skyrme model and despite its
importance. We can refer to an early work on the subject in \cite{newref2}.
Here we shall study analytic results over the properties of such
configurations. Experimentally, \ Baryon-antiBaryon bound states and
resonances do exist \cite{expbab1,expbab2,expbab3}: these should correspond
to Skyrmion-antiSkyrmion bound states. Such bound states are very difficult
to find since the corresponding classical solutions are not static. Indeed,
at a semi-classical level, Skyrmion-antiSkyrmion bound states should look
like time-periodic solutions in which a Skyrmion and an antiSkyrmion moves
periodically around the center of mass of the system. \ These kinds of
time-dependent configurations are difficult to analyze even numerically.

\textbf{3)} \textit{Conductivities}: the analysis of electrons transport
through gauged Skyrmions is a very interesting open issue. At semi-classical
level, one should solve the Dirac equation for the electron in the
background of the gauged Skyrmion and, from the solution of the Dirac
equation, one could compute the conductivity. It would be especially
interesting to be able to describe complex structures assembled from
neutrons and protons interacting with electromagnetic fields (such as slabs
of Baryons interacting with the corresponding Maxwell field). In nuclear
physics and astrophysics these structures are called \textit{nuclear pasta}
and they are very relevant in a huge variety of phenomena (see, for
instance, \cite{nuclearpasta1,nuclearpasta2,nuclearpasta3,nuclearpasta4} and
references therein). On the other hand, there are very few ``first
principles" computations of the transport properties of these complex
structures (see \cite{conductivity} and references therein). At a first
glance, one could think that this kind of complex structure is beyond the
reach of the gauged Skyrme model.

In order to achieve a deeper understanding of the above open issues, it is
mandatory to be able to construct analytic examples of gauged
multi-Skyrmionic configurations.

In \cite%
{canfora2,canfora3,canfora4,canfora4.5,yang1,canfora6,canfora6.5,cantalla4,cantalla5}
a strategy has been developed to generalize the usual spherical hedgehog
ansatz to situations without spherical symmetry both in Skyrme and
Yang-Mills theories (see \cite{canYM1,canYM2,canYM3} and references
therein). Such a framework also allows to analyze configurations living
within a finite region of space.

As far as the three open issues described above are concerned, this tool
(which will be called here ``generalized hedgehog ansatz") gave rise to the
first derivation not only of the critical isospin chemical potential beyond
which the Skyrmion living in the box ceases to exist, but also of the first
explicit Skyrmion-antiSkyrmion bound states. Thus, this approach appears to
be suitable to deal with the problems mentioned previously.

Interestingly enough, the generalized hedgehog ansatz can be adapted to the $%
U(1)$ gauged Skyrme model \cite{Fab1,gaugsk}: it allowed the construction of
two types of gauged solitons. Firstly, gauged Skyrmions living within a
finite volume. Secondly, smooth solutions of the $U(1)$ gauged Skyrme model
whose periodic time-dependence is protected by a topological conservation
law (as they cannot be deformed to static solutions).

Here we demonstrate that by using this strategy it is possible to derive an
explicit expression of the compression modulus. The transport properties of
these gauged Skyrmions can also be analyzed. In this work we also present a
simple estimate of the order of magnitude of the correction to the electron
conductivities due to the interactions of the electrons with the baryonic
environment. As far as transport properties are concerned, we will work at
the level of approximation in which the electrons perceive the gauged
Skyrmions as a classical background. Large \textbf{N} arguments strongly
suggest that this is a very good approximation\footnote{%
In the leading 't Hooft approximation, in meson-Baryon scattering, the heavy
Baryon (the Skyrmion in our case) is unaffected and, basically, only the
meson can react. This is even more so in the electron-Baryon semiclassical
interactions due to the huge mass difference between the Skyrmion and the
electron. In this approximation, electrons perceive the Skyrmions as an
effective medium.} (see for a detailed review chapter 4 and, in particular,
section 4.2 of the classic reference \cite{skyrev0}).

This paper is organized as follows: in the second section the action for the
gauged Skyrme model and our notations will be introduced. In the third
section, the method to deal with Skyrmions at finite density will be
described: as an application, a closed formula for the compression modulus
of Skyrmions living within a cube will be derived. In the fourth section,
the gauged Skyrmions at finite density will be considered. In the fifth
section, the transport properties associated to electrons propagating in the
Baryonic environment corresponding to the finite-density Skyrmions are
analyzed. In section \ref{conclusions}, we draw some concluding ideas.

\section{The $U(1)$ Gauged Skyrme Model}

\label{model}

We consider the $U(1)$ gauged Skyrme model in four dimensions with global $%
SU(2)$ isospin internal symmetry and we will follow closely the conventions
of \cite{Fab1,gaugsk}. The action of the system is
\begin{align}
S& =\int d^{4}x\sqrt{-g}\left[ \frac{K}{2}\left( \frac{1}{2}\mathrm{Tr}%
\left( R^{\mu }R_{\mu }\right) +\frac{\lambda }{16}\mathrm{Tr}\left( G_{\mu
\nu }G^{\mu \nu }\right) \right) -\frac{1}{4}F_{\mu \nu }F^{\mu \nu }\right]
\ ,  \label{sky1} \\
R_{\mu }& =U^{-1}D_{\mu }U\ ,\ \ G_{\mu \nu }=\left[ R_{\mu },R_{\nu }\right]
\ ,\ D_{\mu }=\nabla _{\mu }+\kappa A_{\mu }\left[ t_{3},\ .\ \right] \ ,
\label{sky2} \\
U& \in SU(2)\ ,\ \ R_{\mu }=R_{\mu }^{j}t_{j}\ ,\ \ t_{j}=\mathbbmtt{i}%
\sigma _{j}\ ,  \label{sky2.5}
\end{align}%
where $\sqrt{-g}$ is the (square root of minus) the determinant of the
metric, $F_{\mu \nu }=\partial _{\mu }A_{\nu }-\partial _{\nu }A_{\mu }$ is
the electromagnetic field strength, $\nabla _{\mu }$ is the partial
derivative, the positive parameters $K$ and $\lambda $ are fixed
experimentally, $\kappa $ the coupling for the $U(1)$ field and $\sigma _{j}$
are the Pauli matrices. In our conventions $c=\hbar =\mu _{0}=1$, the
space-time signature is $(-,+,+,+)$ and Greek indices run over space-time.
The stress-energy tensor is
\begin{equation}
T_{\mu \nu }=-\frac{K}{2}\mathrm{Tr}\left[ R_{\mu }R_{\nu }-\frac{1}{2}%
g_{\mu \nu }R^{\alpha }R_{\alpha }\right. \,+\left. \frac{\lambda }{4}\left(
g^{\alpha \beta }G_{\mu \alpha }G_{\nu \beta }-\frac{g_{\mu \nu }}{4}%
G_{\sigma \rho }G^{\sigma \rho }\right) \right] +\bar{T}_{\mu \nu },  \notag
\label{timunu1}
\end{equation}%
with
\begin{equation}
\bar{T}_{\mu \nu }=F_{\mu \alpha }F_{\nu }^{\;\alpha }-\frac{1}{4}F_{\alpha
\beta }F^{\alpha \beta }g_{\mu \nu }.
\end{equation}%
The field equations are
\begin{equation}
D^{\mu }\left( R_{\mu }+\frac{\lambda }{4}\left[ R^{\nu },G_{\mu \nu }\right]
\right) =0\ ,  \label{nonlinearsigma1}
\end{equation}%
\begin{equation}
\nabla _{\mu }F^{\mu \nu }=J^{\nu }\ ,  \label{maxwellskyrme1}
\end{equation}%
where $J^{\nu }$ is the variation of the Skyrme action (the first two terms
in Eq. (\ref{sky1})) with respect to $A_{\nu }$
\begin{equation}
J^{\mu }=\frac{\kappa K}{2}Tr\left[ \widehat{O}R^{\mu }+\frac{\lambda }{4}%
\widehat{O}\left[ R_{\nu },G^{\mu \nu }\right] \right] \ ,  \label{current}
\end{equation}%
where%
\begin{equation*}
\widehat{O}=U^{-1}t_{3}U-t_{3}\ .
\end{equation*}

In the following sections, \textit{gauged Skyrmions} and \textit{gauged
time-crystals} will be terms describing to the two different kinds of gauged
topological solitons appearing as solutions of the coupled system expressed
by Eqs. (\ref{nonlinearsigma1}) and (\ref{maxwellskyrme1}).

The aim of the present work is to show that the Skyrme model and its gauged
version are able to give good predictions for important quantities such as
the compression modulus and the conductivity.

\subsection{Topological charge}

The proper way to define the topological charge in the presence of a minimal
coupling with a $U(1)$ gauge potential has been constructed in \cite{Witten}
(see also the pedagogical analysis in \cite{gaugesky1}):%
\begin{equation}
\begin{split}
W=& \frac{1}{24\pi ^{2}}\int_{\Sigma }\epsilon ^{ijk}Tr\left\{ \left(
U^{-1}\partial _{i}U\right) \left( U^{-1}\partial _{j}U\right) \left(
U^{-1}\partial _{k}U\right) \right. - \\
& \left. \partial _{i}\left[ 3\kappa A_{j}t_{3}\left( U^{-1}\partial
_{k}U+\partial _{k}UU^{-1}\right) \right] \right\} .
\end{split}
\label{new4.1}
\end{equation}

In the literature one usually only considers situations where $\Sigma $ is a
space-like three-dimensional hypersurface. In these situations $W$ is the
Baryon charge. In fact it has been recently shown \cite{Fab1} \cite{gaugsk}
that it is very interesting to also consider cases in which $\Sigma $ is
time-like or light-like. Indeed, (whether $\Sigma $ is light-like, time-like
or space-like) configurations with $W\neq 0$ cannot decay into the trivial
vacuum $U=\mathbb{\mathbf{I}}$. Hence, if one is able to construct
configurations such that $W\neq 0$ along a time-like $\Sigma $, then the
corresponding gauged soliton possesses a topologically protected
time-dependence as it cannot be continuously deformed into static solutions
(since all the static solutions have $W=0$ along a time-like $\Sigma $). The
natural name for these solitons is ``(gauged) time-crystals" \cite%
{Fab1,gaugsk}.

We can adopt the standard parametrization of the $SU(2)$-valued scalar $%
U(x^{\mu }) $
\begin{equation}
U^{\pm 1}(x^{\mu })=Y^{0}(x^{\mu })\mathbb{\mathbf{I}}\pm Y^{i}(x^{\mu
})t_{i}\ ,\ \ \left( Y^{0}\right) ^{2}+Y^{i}Y_{i}=1\,,  \label{standnorm}
\end{equation}%
where $\mathbb{\mathbf{I}}$ is the $2\times 2$ identity and
\begin{align}
Y^{0}& =\cos C\ ,\ Y^{i}=n^{i}\cdot \sin C\ ,  \label{pions1} \\
n^{1}& =\sin F\sin G\ ,\ \ n^{2}=\sin F\cos G\ ,\ \ n^{3}=\cos F\ .
\label{pions2}
\end{align}
with the help of which the standard baryon density (in the absence of a $%
U(1) $ field) reads $\rho _{B}=12\sin ^{2}C\sin F\ dC\wedge dF\wedge dG$. If
we want a non-vanishing topological charge in this setting we have to demand
$dC\wedge dF\wedge dG\neq 0$.

\section{Skyrmions at finite volume}

In the present section, the Skyrmions living within a finite flat box
constructed in \cite{Fab1} will be slightly generalized. These explicit
Skyrmionic configurations allow the explicit computations of the total
energy of the system and, in particular, of its dependence on the Baryon
charge and on the volume. Hence, among other things, one can arrive at a
well-defined closed formula for the compression modulus.

The following anstatz for the representation of the $SU(2)$ group is the
starting point of the analysis
\begin{equation}
G=\frac{q\phi - p\gamma}{2},\; \tan F=\frac{\tan H}{\sin A}, \; \tan C= \tan
A \sqrt{1+\tan ^{2}F}\ ,  \label{pions2.25}
\end{equation}%
where
\begin{equation}
A=\frac{p\gamma +q\phi }{2\,}\ ,\ \ H=H\left( r,z\right) \ ,\ \ p,q\in
\mathbb{N}
\ .  \label{pions2.26}
\end{equation}

Moreover, it can be verified directly that, the topological density $\rho
_{B}$ is non-vanishing. From the standard parametrization of $SU(2)$ \cite%
{Shnir} it follows that
\begin{equation}
0\leq \gamma \leq 4\pi ,\quad 0\leq \phi \leq 2\pi \ ,  \label{domain}
\end{equation}%
while the boundary condition for $H$ will be discussed below; in any case,
its range is in the segment $H \in [0,\frac{\pi}{2}]$, while for $r$ we
assume $0\leq r\leq 2 \pi$. With the parametrization introduced by %
\eqref{pions2.25} and \eqref{pions2.26} the $SU(2)$ field assumes the form
\begin{equation}
U = \pm
\begin{pmatrix}
\cos (H) e^{\frac{1}{2} i (p \gamma+q \phi )} & \sin (H) e^{\frac{1}{2} i (p
\gamma -q \phi )} \\
-\sin (H) e^{-\frac{1}{2} i (p \gamma -q \phi )} & \cos (H) e^{-\frac{1}{2}
i (p\gamma +q \phi )}%
\end{pmatrix}%
.
\end{equation}
Hereafter, we just consider the plus expression for $U$ throughout all the
range of the variables $\gamma$ and $\phi$, which makes it a continuous
function of the latter.

\subsection{Skyrmions in a rectangular cuboid}

We can extend the results presented in \cite{Fab1} by considering a cuboid
with three different sizes along the three axis instead of a cube. Thus, we
will use three - different in principle - fundamental lengths characterizing
each direction, $l_{1}$, $l_{2}$ and $l_{3}$, inside the metric.

The corresponding line element is
\begin{equation}
ds^{2}=-dz^{2}+l_{1}^{2}dr^{2}+l_{2}^{2}d\gamma ^{2}+l_{3}^{2}d\phi ^{2}\ .
\label{line3l}
\end{equation}%
The profile function that we consider depends only on one variable\footnote{%
On the other hand, when the coupling with Maxwell field is neglected, the
profile can depend on time as well. In this case, one gets an effective
sine-Gordon theory for the profile $H(t,r)$ \cite{Fab1}.}, $H=H(r)$. We note
that in this section we do not take into account the effects of an
electromagnetic field, hence we have $A_{\mu }=0$ in the relations of the
previous sections.

Under the aforementioned conditions the profile equation reduces to
\begin{equation}  \label{profnoA}
H^{\prime \prime }= \frac{\lambda l_1^2 p^2 q^2}{4 \left(l_2^2 \left(4
l_3^2+\lambda q^2\right)+\lambda l_3^2 p^2\right)} \sin (4 H).
\end{equation}
It is impressive that such a system, in flat space, can lead to an
integrable equation for the profile. This is owed to the existence
of a first integral of \eqref{profnoA} that is given by
\begin{equation}  \label{intofmol3}
(H^{\prime})^2 \left(l_2^2 \left(4 l_3^2+\lambda q^2\right)+\lambda l_3^2
p^2\right)+\frac{\lambda l_1^2 p^2 q^2}{8} \cos (4 H) = I_0 .
\end{equation}
The above relation can be written as
\begin{equation}  \label{firstint2}
(\tilde{H}^{\prime})^2 - k \sin (\tilde{H})^2 = \tilde{I}_0,
\end{equation}
where
\begin{equation}
\tilde{H}= 2 H, \quad k = \frac{\lambda l_1^2 p^2 q^2}{l_2^2 \left(4
l_3^2+\lambda q^2\right)+\lambda l_3^2 p^2}, \quad \tilde{I}_0= \frac{8
I_0-\lambda l_1^2 p^2 q^2}{8 l_2^2 l_3^2+2 \lambda l_2^2 q^2+2 \lambda l_3^2
p^2}.
\end{equation}
Subsequently, we can bring \eqref{firstint2} into the form
\begin{equation}  \label{firstint3}
\frac{d \tilde{H}}{d r} = \pm \sqrt{\tilde{I}_0} \left(1 - \tilde{k}(\sin
\tilde{H})^2 \right)^{\frac{1}{2}}
\end{equation}
where we have set $\tilde{k} = - k/\tilde{I}_0$. The last expression leads
to
\begin{equation}  \label{finalsolH}
\sqrt{\tilde{I}_0} \int_0^r d \bar{r} = \pm \int_0^{\tilde{H}} \left(1 -
\tilde{k}(\sin \bar{H})^2 \right)^{-\frac{1}{2}} d \bar{H},
\end{equation}
where we have introduced the bars in order to distinguish the variables that
are integrated from the $r$ and $\tilde{H}(r)$ which are the boundaries of
the two integrals. Of course we consider $\tilde{I}_0 > 0$. As a starting
point for the integration we take $r=0$, $\tilde{H}(0)=0=H(0)$, although we
could also set $r=0$, $\tilde{H}=\pi$ ($H(0)=\frac{\pi}{2}$). The difference
between the two boundary choices is just in the sign of the topological
charge. These boundary values, for $H$ and those that we have seen in %
\eqref{domain} for $\gamma$ and $\phi$ lead to a topological charge $W= p q$
in \eqref{new4.1} (for $A_\mu =0$).

In the right hand side of \eqref{finalsolH} we recognize the
incomplete elliptic integral defined as
\begin{equation}
F(\tilde{H}|\tilde{k}) = \int_0^{\tilde{H}} \left(1 - \tilde{k}(\sin \bar{H}%
)^2 \right)^{-\frac{1}{2}} d \bar{H} .
\end{equation}
The solution to the differential equation \eqref{firstint3} is just the
inverse of this function, which is called the Jacobi amplitude $\mathrm{am}%
=F^{-1}(\tilde{H}|\tilde{k})$. So, in terms of our original equation %
\eqref{profnoA} the solution reads
\begin{equation}  \label{soltodifeq}
H(r) = \pm \frac{1}{2} \mathrm{am}(\tilde{I}_0^{1/2} r|\tilde{k}).
\end{equation}
Finally, by considering the positive branch, the value of the constant of
integration $\tilde{I}_0$ is governed by the boundary condition $H(2\pi)=%
\frac{\pi}{2}$.

\begin{figure}[h]
\centering
\includegraphics[width=.40\textwidth]{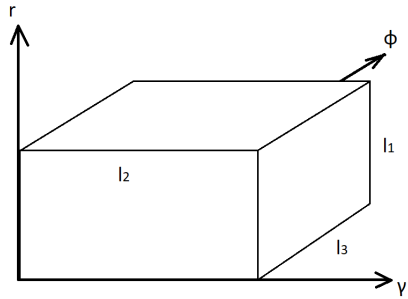}
\caption{The finite box of the Skyrmionic system.}
\label{Fig0}
\end{figure}

In the special case when $l_1=l_2=l_3=l$ we obtain the particular case which
was studied in \cite{Fab1}. Here, we give emphasis to this general case and,
especially, we want to study the most energetically convenient
configurations and the way in which they are affected by the anisotropy in
the three spatial directions. In Fig. \ref{Fig0} we see a schematic
representation of the finite box we are considering for this Skyrmionic
configuration with a baryon number $B =p q$.

The physical configuration that we try to reproduce with this model
is the structure of matter in nuclear pasta. The latter is a dense form of
matter that is encountered inside the crusts of neutron stars. Thus, we make
this \textquotedblleft crude" (but analytic in its results) model trying to
imitate with these $p$ and $q$ Skyrmionic layers a particular form of this
matter that is encountered in nature. The dimensions of the configuration
are governed by the three numbers $l_{1}$, $l_{2}$ and $l_{3}$. Of course we
do not expect the binding energies of such a configuration to be at the same
level with those produced by the usual spherically symmetric ansatz. This is
something that we examine thoroughly in the next section.

\subsubsection{The energy function}

We proceed to study the energy function for the solution that we previously
introduced. The constant of motion $I_0$ in \eqref{intofmol3} can be
expressed in terms of the other constants of the model if we consider the
boundary values $H(0)=0$ and $H(2\pi)=\pi/2$. By solving \eqref{intofmol3}
with respect to $H^{\prime }$ and integrating the resulting relation with
respect to $r$ we obtain
\begin{equation}  \label{intofint}
2\sqrt{2} \int_{a}^{b} \left(\frac{l_2^2 \left(4 l_3^2+q^2\right)+l_3^2 p^2}{%
8 I_0-l_1^2 p^2 q^2 \cos (4 H)}\right)^{1/2} dH = \int_{0}^{2\pi} dr
\end{equation}
which leads to
\begin{equation}  \label{l1tox}
l_1 = \frac{x \mathrm{K}\left(-x^2\right) \sqrt{l_2^2 \left(4
l_3^2+q^2\right)+l_3^2 p^2}}{\pi p q},
\end{equation}
where $\mathrm{K}$ is the complete elliptic integral of the first kind and $%
x $ is related to $I_0$ through
\begin{equation}  \label{I0tox}
I_0 = \frac{l_1^2 p^2 q^2 \left(x^2+2\right)}{8 x^2}.
\end{equation}

\begin{figure}[tbp]
\centering
\includegraphics[width=.40\textwidth]{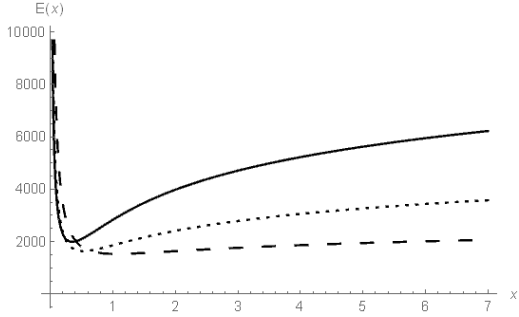}
\caption{The plots of $E(x)$ (in MeV) for three sets of values: (a) $p=q=3$,
$l_2=l_3=1$ fm (dashed line), (b) $p=q=3$, $l_2=l_3=2$ fm (dotted line) and
(c) $p=q=3$, $l_2=l_3=3$ fm (continuous line). The minimum of the energy
corresponds to $l_1=0.227$ fm, $l_1=0.323$ fm and $l_1=0.42$ fm
respectively. }
\label{Fig1}
\end{figure}

The pure time component of the energy momentum tensor in our case is
\begin{equation}
T_{00} = \frac{K}{8 V^2} \left[\left(l_2^2 \left(4 l_3^2+\lambda
q^2\right)+\lambda l_3^2 p^2\right) H^{\prime 2 }+ \frac{\lambda l_1^2 p^2
q^2}{4}\sin ^2(2 H) + V^2 \left(\frac{p^2}{l_2^2}+\frac{q^2}{l_3^2}\right)%
\right].
\end{equation}
As a result we can calculate the energy from the expression
\begin{equation}
E= \int_{\Sigma} \sqrt{-{}^{(3)}g} T_{00}d^3 x = 8\pi^2 V \int_0^{\frac{\pi}{%
2}} \frac{T_{00}}{H^{\prime }} dH .
\end{equation}
We can write the integrand as a pure function of $H$ with the help of %
\eqref{intofmol3} and obtain - in principle - the energy as a function of
the $l_i$' s, $p$ and $q$. However, due to the fact that relation %
\eqref{l1tox} cannot be straightforwardly inverted so as to substitute $I_0$
as a function of $l_1$ (through \eqref{l1tox} and \eqref{I0tox}) we choose
to express the energy function in terms of $x$ instead of $l_1$. In what
follows, we assume the values $K=2$ and $\lambda=1$ for the coupling
constants \cite{skyrev1}, so that lengths are measured in fm and the energy
in MeV. In this manner we get
\begin{equation}  \label{energyfull}
E(x,l_2,l_3,p,q) = \frac{\pi ^2 p q \sqrt{l_2^2 \left(4
l_3^2+q^2\right)+l_3^2 p^2}}{l_2 l_3} \frac{K(-x^2) \left(\frac{4 l_2^2 x^2
K(-x^2)}{p^2}-\frac{\mathrm{K}(-x^2) \left(q^2-4 l_3^2 x^2\right)}{q^2}+2
\mathcal{E}(-x^2)\right)}{x |K(-x^2)|},
\end{equation}
where $\mathcal{E}$ is the complete elliptic integral of the second kind.
The $x$, as we discussed, is linked - with the help of the boundary
conditions of the problem - through \eqref{l1tox} to $l_1$. If we fix all
variables apart from $x$ and plot the energy as a function of the latter we
get what we see in Fig. \ref{Fig1}. In this graph, we observe that the
minimum of the energy is ``moving" to smaller values of $x$ as the box is
being enlarged in the two directions of $l_2$ and $l_3$. However, we have to
keep in mind that the other of the lengths, namely $l_1$, depends also on
the values of $l_2$ and $l_3$ through \eqref{l1tox}. For the particular set
of values used in the figure we can see that as $l_2$ and $l_3$ rise, $l_1$
is also relocated to larger values. In the next section we study more
thoroughly the function $E(x,l_2,l_3,p,q)$ and its derivatives near the
values that correspond to the most energetically convenient configurations.

\subsubsection{The energy as a function of the three $l_i$'s}

Let us see how the energy behaves in terms of the three fundamental lengths $%
l_1$, $l_2$ and $l_3$ under the condition that we fix $p$ and $q$ to
specific values. In the table \ref{tab1} we can observe the location of the
minimum of the energy for specific values of $p$ and $q$.

\begin{table}[tbp]
\begin{center}
\begin{tabular}{|c|c|c|c|c|c|}
\hline
$E_{min}$ (MeV) & $p$ & $q$ & $l_1$ (fm) & $l_2$ (fm) & $l_3$ (fm) \\ \hline
167 & 1 & 1 & 0.251 & 0.413 & 0.413 \\
334 & 1 & 2 & 0.251 & 0.413 & 0.826 \\
669 & 2 & 2 & 0.251 & 0.826 & 0.826 \\
835638 & 100 & 50 & 0.251 & 41.306 & 20.653 \\
835638 & 50 & 100 & 0.251 & 20.653 & 41.306 \\ \hline
\end{tabular}%
\end{center}
\caption{Minimum of the energy for values of $p$ and $q$.}
\label{tab1}
\end{table}

First, we have to note that the interchange of $p$ and $q$ makes no
significant difference, so weather you take $p=100$ and $q=50$ or $p=50$ and
$q=100$, the only thing that happens is that the values of the corresponding
lengths $l_{2}$ and $l_{3}$ are also interchanged. However, the arithmetic
value that the energy assumes remains the same. Another thing that we have
to notice is that, if we calculate the percentage difference of the minimum
of the energy from the topological bound $E_{0}=12\pi ^{2}|B|=12\pi ^{2}pq$;
in all cases we get $\Delta (\%)=\frac{E-E_{0}}{E_{0}}(\%)=41.11\%$. Thus,
we see that the minimum of the energy $E(l_{1},l_{2},l_{3})$ has a fixed
deviation from the Bogomol'nyi bound irrespectively of the $p$, $q$
configuration. We also observe that this most energetically convenient
situation arises when the box has convenient lengths. In particular we see
that the relation $\frac{l_{2}}{l_{3}}=\frac{p}{q}$ is satisfied in all
cases, while $l_{1}$ remains fixed in a single \textquotedblleft optimal"
value. By comparing with the usual spherically symmetry Skyrmionic
configuration in an infinite volume, this higher deviation from the
Bogomol'nyi bound may be anticipated due to the ``compression" of the system
into a finite volume.

\begin{figure}[h]
\centering
\hspace{8mm}
\subfloat[][Derivative of the energy with respect to
$l_1$]{\includegraphics[width=.40 \textwidth]{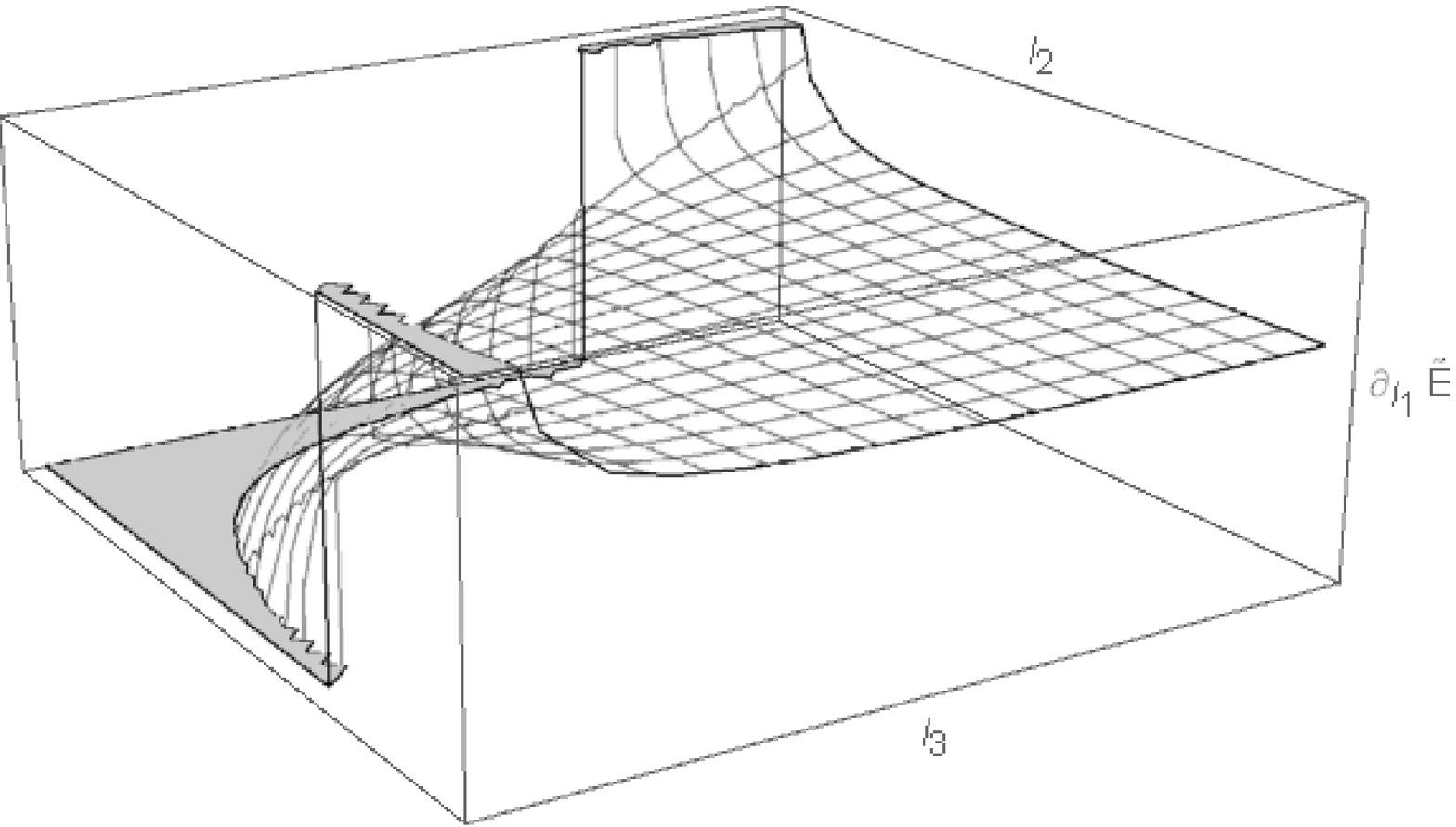}} \hspace{0mm}
\subfloat[][Derivative of the energy with respect to
$l_2$]{\includegraphics[width=.41 \textwidth]{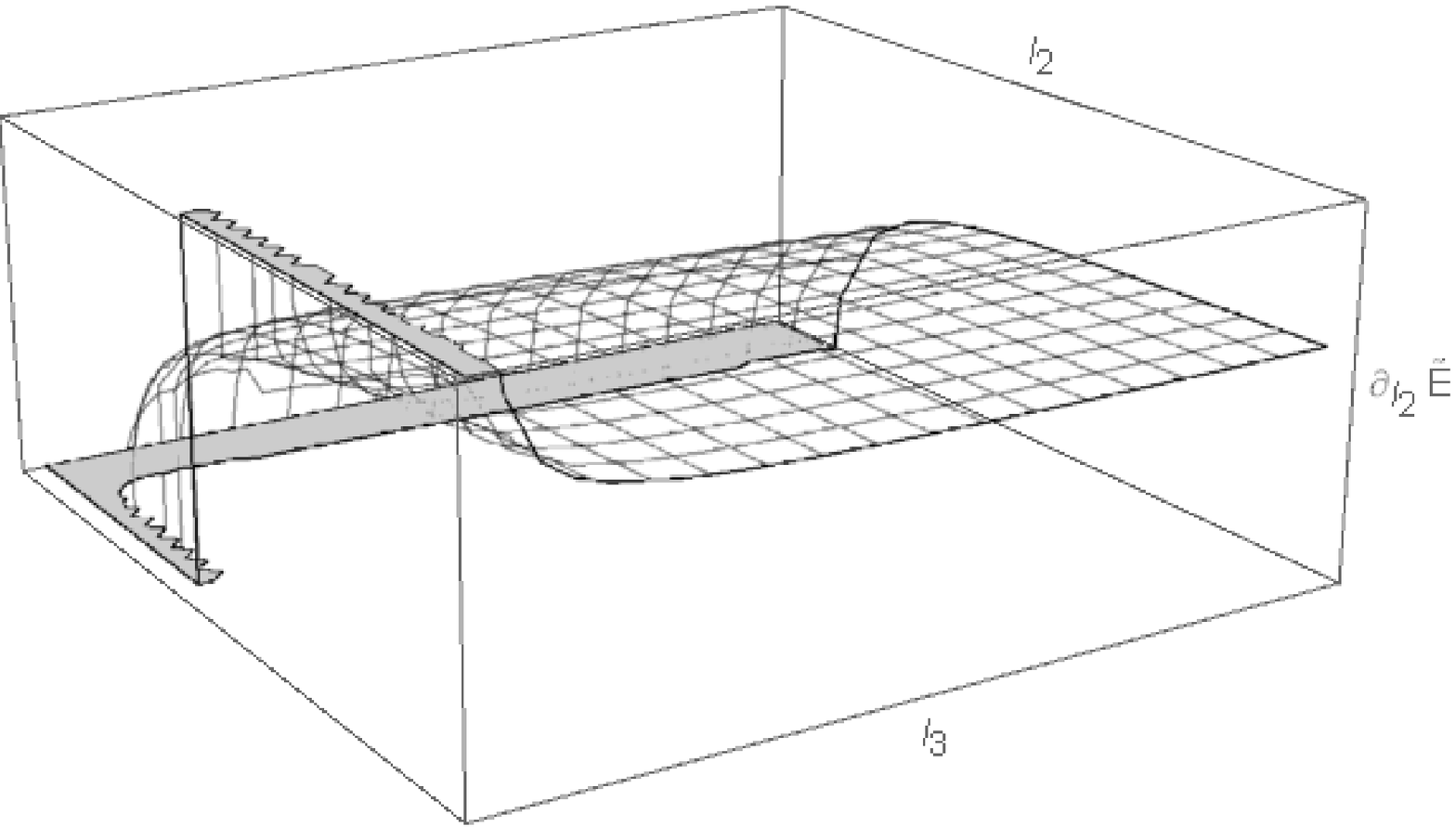}} \hspace{0mm}
\subfloat[][Derivative of the energy with respect to
$l_3$]{\includegraphics[width=.40\textwidth]{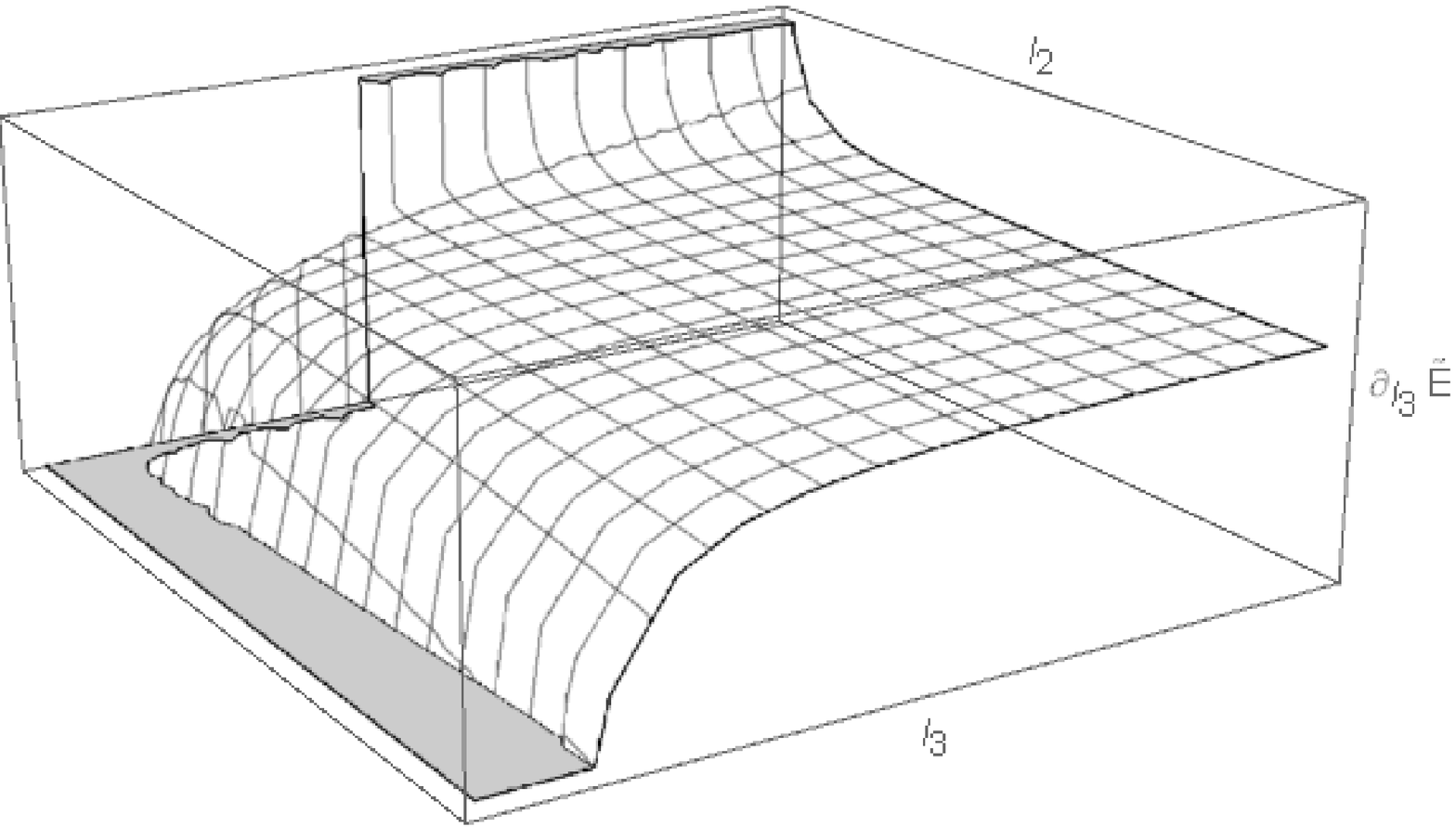}}
\caption{Derivative of the energy, in terms of the basic dimensions of the
Skyrmionic box, near its minimum value. The behaviour of the three $\frac{%
\partial E}{\partial l_i}$ is the same irrespectively of $p$ and $q$. The
only thing that changes is the scaling of the figures since $l_2$ and $l_3$
and $\frac{\partial E}{\partial l_i}$ assume larger values as $p$ and $q$
increase.}
\label{Fig2}
\end{figure}

\begin{figure}[tbp]
\centering
\hspace{8mm}
\subfloat[][$\frac{\partial E}{\partial l_i}$ for $p=1$,
$q=2$]{\includegraphics[width=.40 \textwidth]{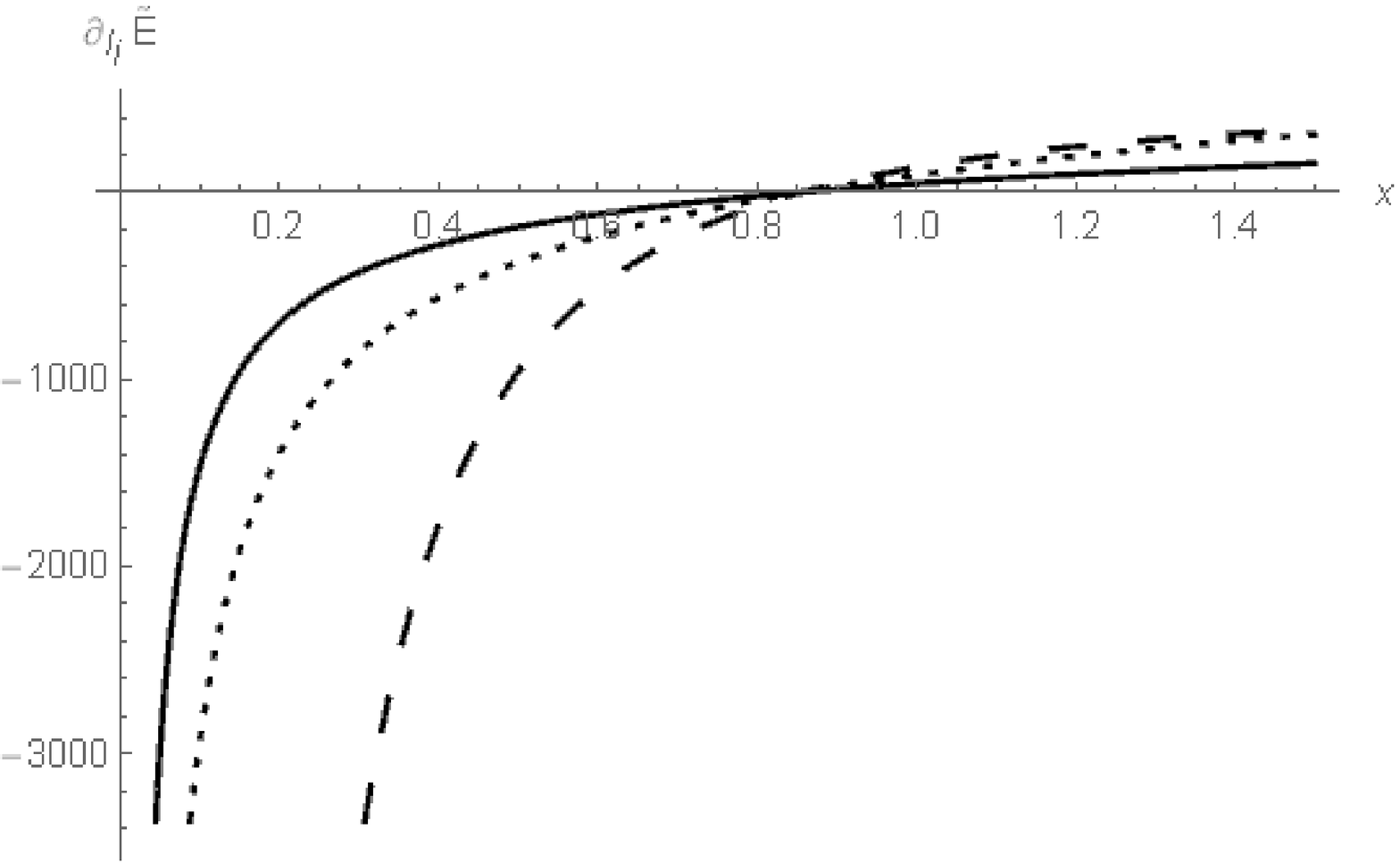}} \hspace{0mm}
\subfloat[][$\frac{\partial E}{\partial l_i}$ for $p=2$,
$q=2$]{\includegraphics[width=.41 \textwidth]{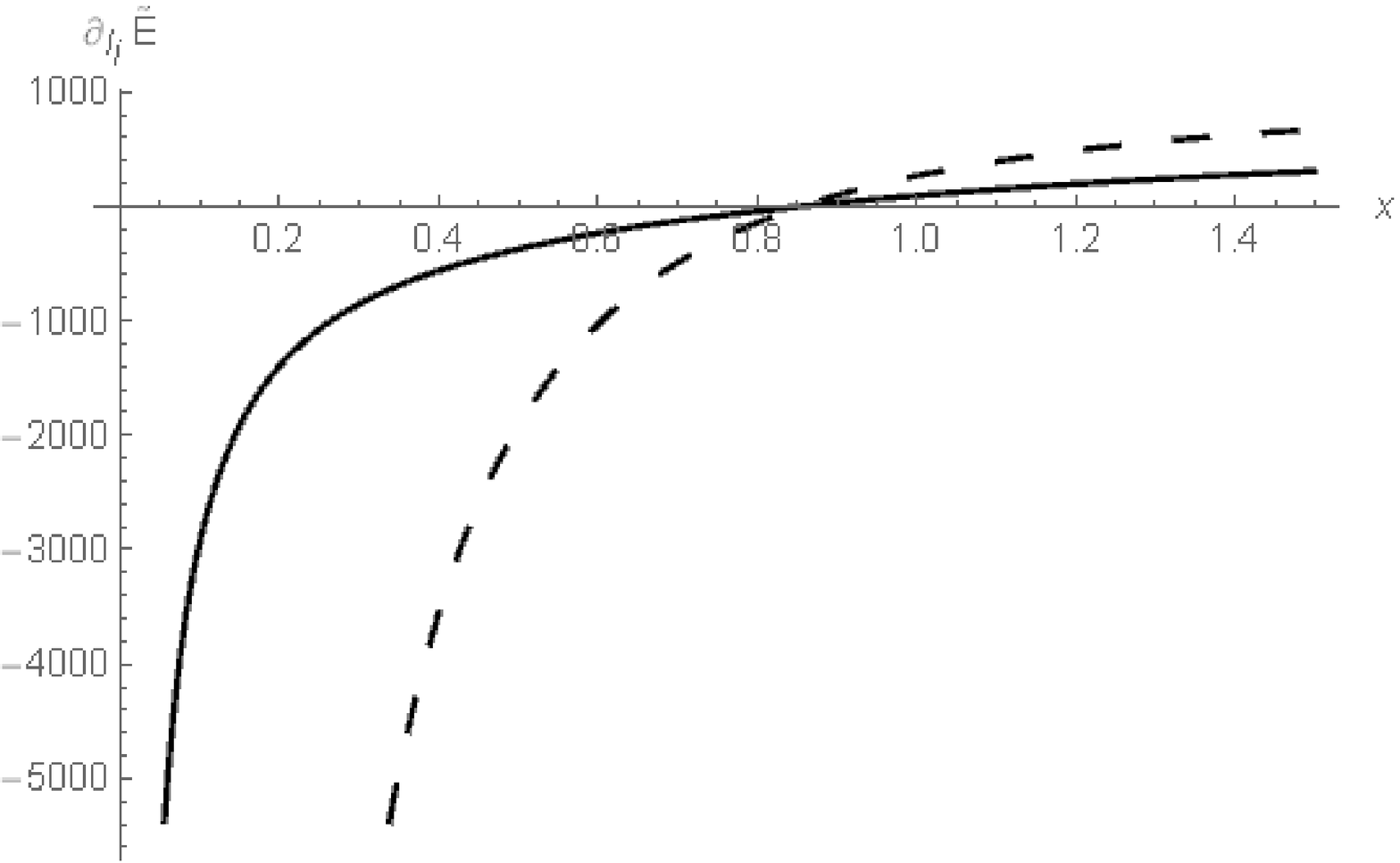}} \hspace{0mm}
\subfloat[][$\frac{\partial E}{\partial l_i}$ for $p=100$,
$q=50$]{\includegraphics[width=.40\textwidth]{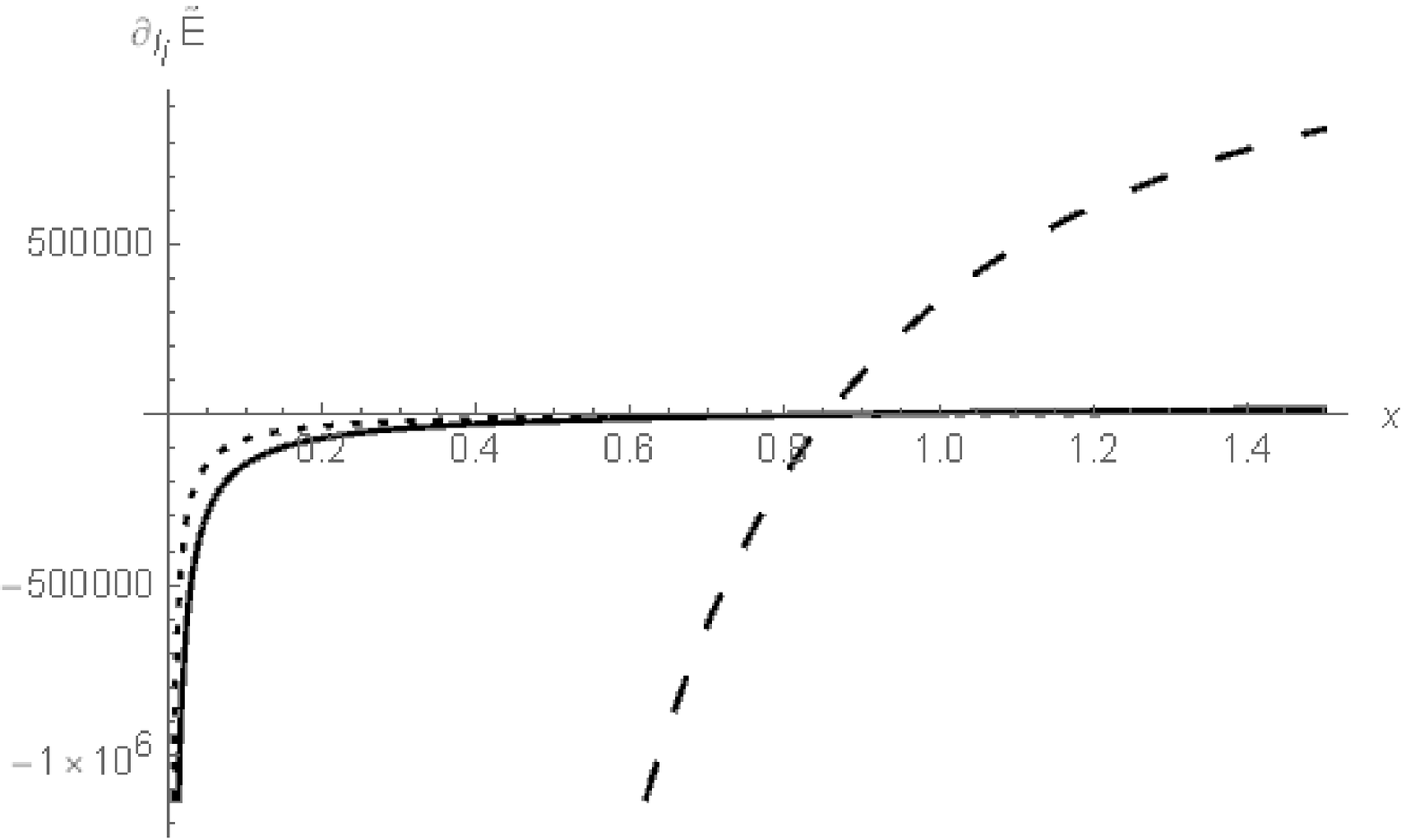}}
\caption{Derivative of the energy with respect to the $l_i$'s given as
function of $x$. In every case the dashed line corresponds to $\frac{%
\partial E}{\partial l_1}$, the dotted to $\frac{\partial E}{\partial l_2}$
and the continuous line to $\frac{\partial E}{\partial l_3}$. Lengths are
measured in fm and the energy in MeV.}
\label{Fig3}
\end{figure}

\begin{figure}[tbp]
\centering
\includegraphics[width=.40\textwidth]{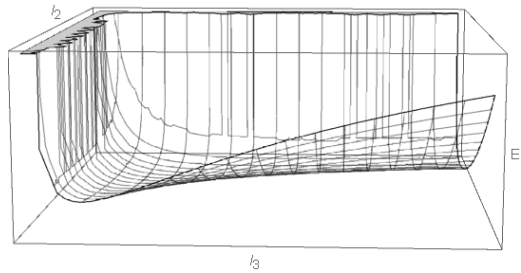}
\caption{Plot of the energy $E$ in the $l_{2}-l_{3}$ plane when $l_{1}$
takes the value that corresponds to the minimum of $E$.}
\label{Fig4}
\end{figure}

It is also interesting to study the first derivatives of the energy with
respect to the three lengths of the box. To this end, and since we have $E$
in terms of $x$ which also involves $l_{1}$, $l_{2}$ and $l_{3}$ we need to
write
\begin{equation}
\begin{split}
dE(x,l_{2},l_{3})& =\frac{\partial E}{\partial x}dx+\frac{\partial E}{%
\partial l_{2}}dl_{2}+\frac{\partial E}{\partial l_{3}}dl_{3} \\
& =\frac{\partial E}{\partial x}\frac{\partial x}{\partial l_{1}}%
dl_{1}+\left( \frac{\partial E}{\partial x}\frac{\partial x}{\partial l_{2}}+%
\frac{\partial E}{\partial l_{2}}\right) dl_{2}+\left( \frac{\partial E}{%
\partial x}\frac{\partial x}{\partial l_{3}}+\frac{\partial E}{\partial l_{3}%
}\right) \\
& =d\tilde{E}(l_{1},l_{2},l_{3}).
\end{split}%
\end{equation}%
In Fig. \ref{Fig2} we can see the general behavior of three $\frac{\partial
\tilde{E}}{\partial l_{i}}$ for fixed $l_{1}=0.251$ in terms of $l_{2}$ and $%
l_{3}$ near the values where the energy assumes its minimum. On the other
hand, in Fig. \ref{Fig3} we plot the derivatives of the energy with respect
to $x$ after fixing $l_{2}$ and $l_{3}$ to their minimum value for various $%
p $, $q$ configurations. We can see that $\frac{\partial E}{\partial l_{2}}$
and $\frac{\partial E}{\partial l_{3}}$ are indistinguishable when $p=q$. On
the other hand if $q>p$ the $\frac{\partial E}{\partial l_{3}}$ line runs
closer to the vertical axis than $\frac{\partial E}{\partial l_{2}}$ and
vice versa when $p>q$. Finally, before proceeding to study the energy as a
function of $p$ and $q$, we give in Fig. \ref{Fig4} its graph in terms of $%
l_{2}$ and $l_{3}$ when $l_{1}$ assumes the value that corresponds to the
minimum of the energy.

\subsection{The energy of the symmetric configuration}

Due to using \eqref{l1tox} in the previous section so as to write the energy
as a function of $x$, $l_{2}$ and $l_{3}$, it is not straightforward from
that expression to derive what happens in the case where one considers a
symmetric box $l_{1}=l_{2}=l_{3}=l$. In this section we treat this situation
from the very beginning by setting all fundamental lengths as equal in Eq. %
\eqref{intofmol3}. We have to note that throughout this section we also make
use of the system of units $K=2$, $\lambda =1$. The expression relative to %
\eqref{intofint}, from the resulting integral of motion, leads to
\begin{equation}
l=\frac{\sqrt{\pi ^{2}p^{2}q^{2}-x^{2}\mathrm{K}\left( -x^{2}\right)
^{2}\left( p^{2}+q^{2}\right) }}{2x\mathrm{K}\left( -x^{2}\right) },
\label{ltox}
\end{equation}%
where $x$ is defined as in the previous section by relation \eqref{I0tox},
with $l_{1}=l$. By following the exact same steps as before we are led to
the following expression for the energy
\begin{equation}
E_{c}(x,p,q)=\frac{2\pi ^{3}\left( 2p^{2}q^{2}\mathrm{K}\left( -x^{2}\right)
\mathcal{E}\left( -x^{2}\right) -\mathrm{K}\left( -x^{2}\right) ^{2}\left(
p^{4}x^{2}+p^{2}q^{2}\left( 2x^{2}+1\right) +q^{4}x^{2}\right) +\pi
^{2}p^{2}q^{2}\left( p^{2}+q^{2}\right) \right) }{x^{2}\mathrm{K}\left(
-x^{2}\right) \sqrt{\frac{\pi ^{2}p^{2}q^{2}}{x^{2}}-\mathrm{K}\left(
-x^{2}\right) ^{2}\left( p^{2}+q^{2}\right) }}.  \label{encube}
\end{equation}

\begin{figure}[tbp]
\centering
\includegraphics[width=.40\textwidth]{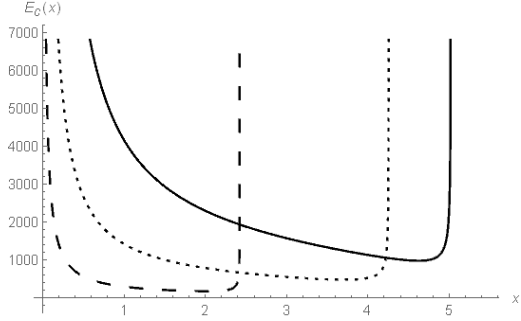}
\caption{Plot of the energy of the cubic configuration $E_c$ with respect to
$x$. The dashed line corresponds to $p=q=1$, the dotted to $p=2$, $q=1$ and
the continuous line to $p=3$, $q=1$. The minimum of the energy in terms of
the size of the cube $l$ is: $l=0.322$, $l=0.369$ and $l=0.385$
respectively. }
\label{Fig4new}
\end{figure}

It is easy to note that the energy is symmetric under the mirror change $%
p\leftrightarrow q$. We verify that the for a bigger baryon number, the most
optimal configuration corresponds also to a larger box. In Fig. \ref{Fig4new}
we can see the plot of the energy with respect to various configurations
demonstrating the aforementioned fact. The second thing that we can note is
that the deviation $\Delta= \frac{E-E_0}{E_0}$ from saturating the bound
also increases for larger baryonic configurations. In table \ref{tab2} we
provide some basic examples. Surprisingly we can see that the configuration $%
p=q=2$ is slightly more convenient than the one corresponding to $p=2$, $q=1$%
. As long as we know, this is the only case where this is happening. In
general it can be seen that the $p=q$ construction requires more energy than
the $p$, $q-1$, with an exception in the $p=q=2$ case.

\begin{table}[tbp]
\begin{center}
\begin{tabular}{|c|c|c|c|}
\hline
$p$ & $q$ & $l$ (fm) & $\Delta(\%)$ \\ \hline
1 & 1 & 0.322 & 53 \\
2 & 1 & 0.369 & 105 \\
3 & 1 & 0.385 & 177 \\
2 & 2 & 0.463 & 104 \\
3 & 2 & 0.505 & 138 \\
3 & 3 & 0.571 & 148 \\ \hline
\end{tabular}%
\end{center}
\caption{Deviation from the topological bound for several values of $p$ and $%
q$.}
\label{tab2}
\end{table}

\subsection{The compression modulus for the rectangular box}

From the technical point of view, it is worth to emphasize here that the
very notion of compression modulus would require to put the Skyrmions within
a finite flat box of volume $V$: then the compression modulus is related to
the second derivative of the total energy of the system with respect to $V$.
As it has been already mentioned, this requires to generalize the hedgehog
ansatz to situations without spherical symmetry. On the other hand, if one
insists in defining the compression modulus for the spherical hedgehog, it
becomes a rather subtle issue (see the nice analysis in \cite{Adam}) how to
define the derivative of the energy with respect to the volume. Here we are
using the generalized hedgehog ansatz \cite{Fab1,gaugsk}\ which is well
suited to deal with situations without spherical symmetry. In this way we
can analyze Skyrmions living within a region of flat space-time of finite
spatial volume avoiding all the subtleties mentioned above. In particular,
in the present case the "derivative with respect to the volume" means,
literally, the derivative (of the total energy of the system) with respect
to the spatial volume of the region in which the Skyrmions are living.

As we obtained the general behavior of the three $\frac{\partial E}{\partial
l_{i}}$ functions in the previous sub-sections, we are also able to derive
an analytic expression of the compression modulus \cite{Brown,Co}
\begin{equation*}
\mathcal{K}=\frac{9V}{B\beta }\approx 210\pm 30MeV
\end{equation*}%
where $\beta =-\frac{1}{V}\frac{\partial V}{\partial P}$ is the
compressibility. By using $P=\frac{dE}{dV}$ we acquire
\begin{equation}
\mathcal{K}=-\frac{9V^{2}}{B}\frac{d^{2}E}{dV^{2}},  \label{compmod}
\end{equation}%
where $B$ is the baryon charge and $V$ the finite volume in which we confine
the system; in our case this volume is $V=16\pi ^{3}l_{1}l_{2}l_{3}$. The
difference in the sign of \eqref{compmod} in comparison to other expressions
in the literature \cite{Blaizot} is owed to the metric signature that we
follow here and which affects the derivation of $E$ from $T_{00}$. In order
to express the energy that we obtain from \eqref{energyfull} as a function
of the volume, we introduce the following reparametrization of the $l_{i}$'s
into three new variables
\begin{equation}
l_{1}=c_{1}\left( \frac{V}{16\pi ^{3}}\right) ^{1/3},\quad l_{2}=c_{2}\left(
\frac{V}{16\pi ^{3}}\right) ^{1/3},\quad \text{and}\quad l_{3}=\frac{1}{%
c_{1}c_{2}}\left( \frac{V}{16\pi ^{3}}\right) ^{1/3},
\end{equation}%
so that $l_{1}l_{2}l_{3}=\frac{V}{16\pi ^{3}}$. We can substitute the above
expressions into both \eqref{l1tox} and \eqref{energyfull}. By solving the
first with respect to $V$ and substituting to the second we obtain the
energy as a pure function of $x$ which is associated through \eqref{l1tox}
with the volume $V$. We can thus calculate the first and second derivatives
of the energy with respect to the volume by just taking $\frac{dE}{dV}%
=\left( \frac{dV}{dx}\right) ^{-1}\frac{dE}{dx}$ and $\frac{d^{2}E}{dV^{2}}%
=\left( \frac{dV}{dx}\right) ^{-1}\frac{d}{dx}\left[ \left( \frac{dV}{dx}%
\right) ^{-1}\frac{dE}{dx}\right] $.

\begin{figure}[h]
\centering
\hspace{0mm}
\subfloat[][Pressure
$P(V)$]{\includegraphics[width=4cm,height=4cm]{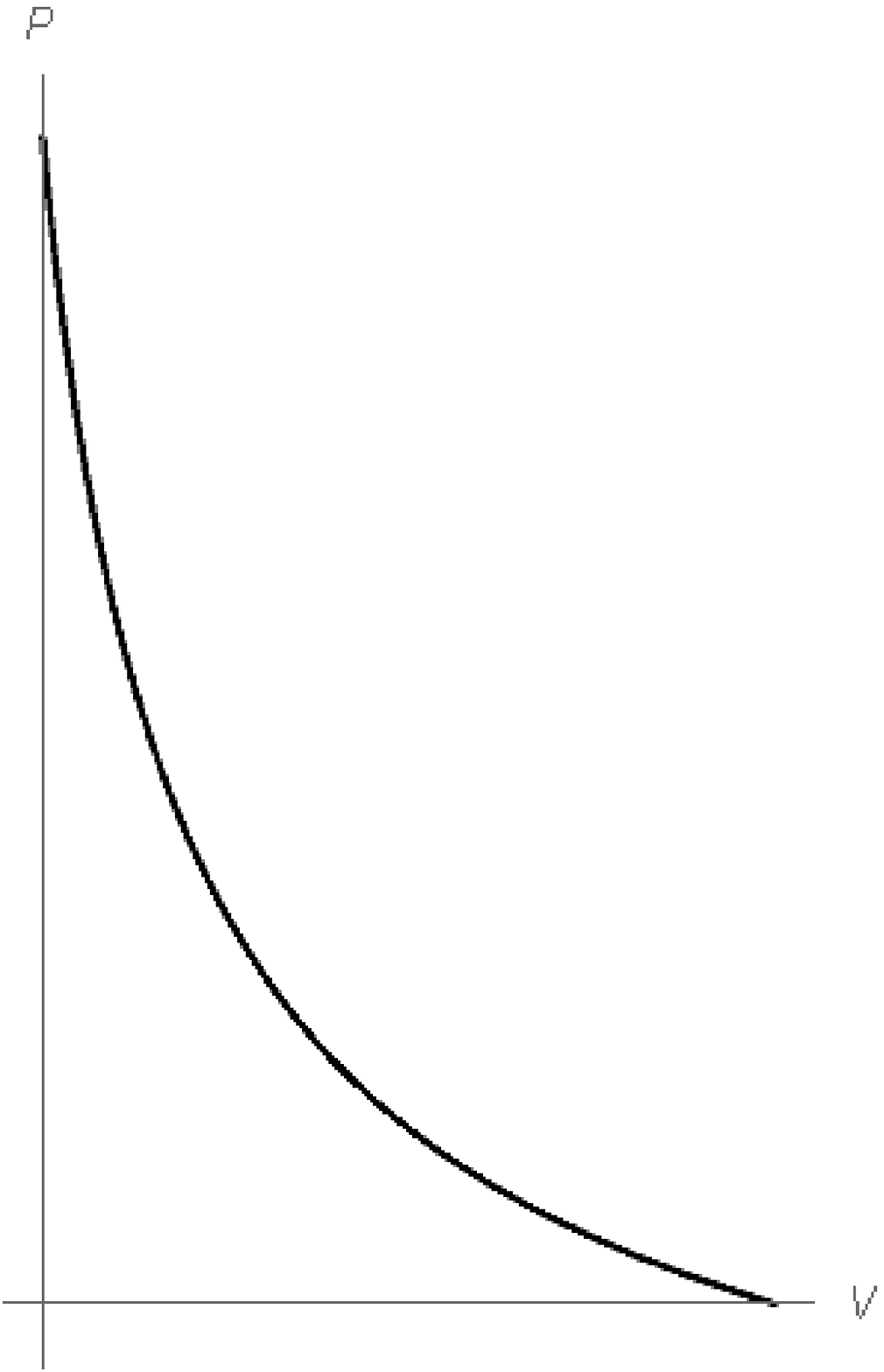}} \hspace{8mm}
\subfloat[][Compression modulus $\mathcal{K}(V)$]{\includegraphics[width=.40
\textwidth]{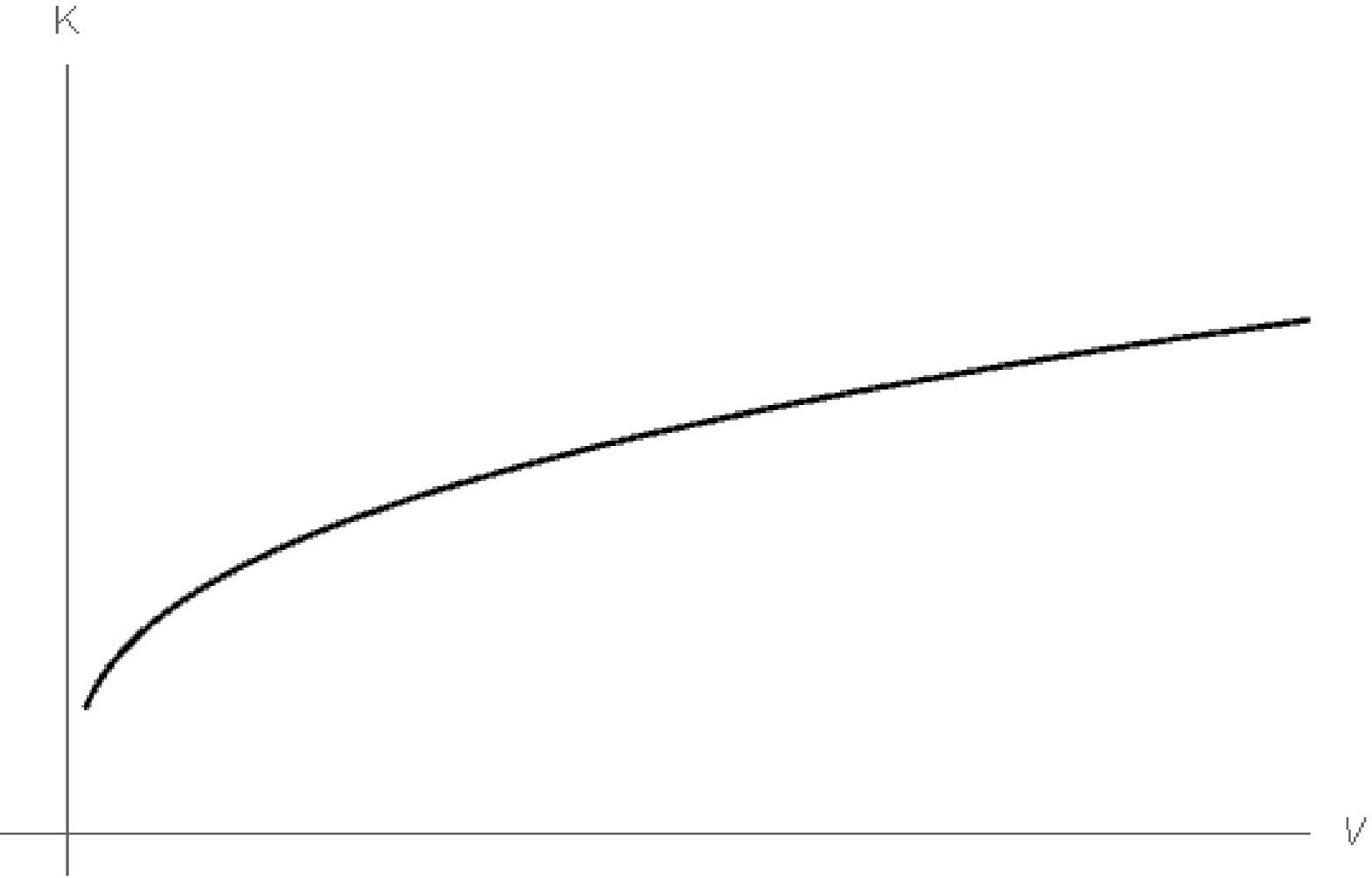}} \hspace{0mm}
\subfloat[][Energy
$E(V)$]{\includegraphics[width=.40\textwidth]{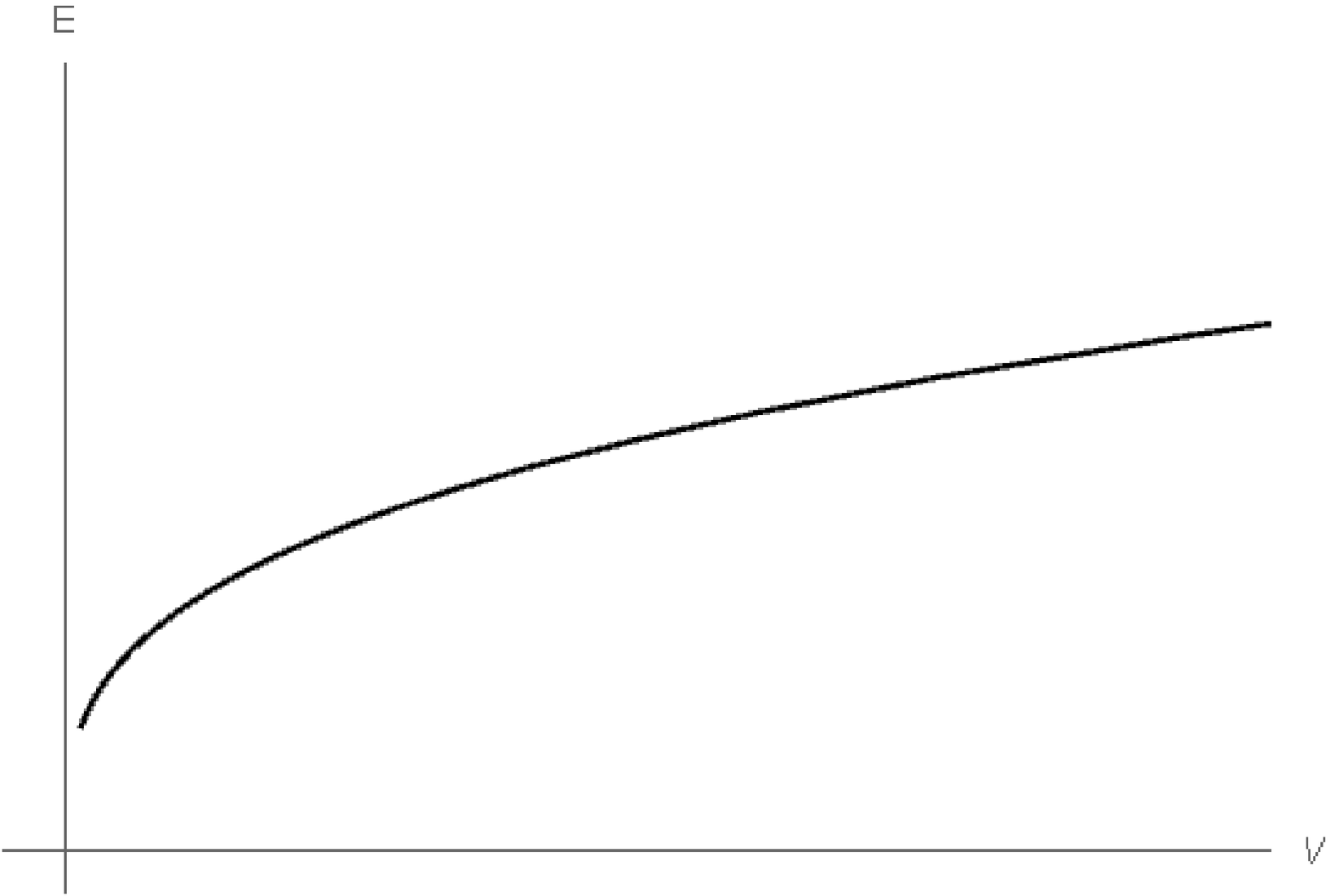}}
\caption{Parametric plots of the pressure $P$, the compression modulus $%
\mathcal{K}$ and the energy $E$ with respect to the volume. The plots
correspond to the same parameters but for different ranges of the volume.}
\label{Fig5}
\end{figure}

The first derivative of $E(V)$ with respect to the volume defines the
pressure of the system, i.e. $P=\frac{dE}{dV}$. In Fig. \ref{Fig5} we see
the graphs of the pressure the compression modulus and the energy with
respect to the volume for specific regions of the variable $V$. Due to the
complicated nature of the relation between $x$ and $V$ it is not easy to put
in this parametric plot the behavior of $P$ and $E$ near the region where $%
V\rightarrow 0$. However, one can calculate through the relations that as
one shrinks the volume to zero, the pressure suddenly falls and changes sign
becoming negative. The same happens to the compression modulus $\mathcal{K}$
as well, for even smaller values of $V$, while the energy remains positive
for all $V$. Unfortunately the expressions are too cumbersome to present
them analytically in this work, but the graphs in Fig. \ref{Fig5}
demonstrate the general behavior. In the case of a finite cube with $%
l_{1}=l_{2}=l_{3}$ the situation is a lot simpler as we can see in the
following section.

\subsubsection{Compression modulus in the symmetric case}

The most natural case corresponds to choose $l_{1}=l_{2}=l_{3}=l$. In this
way, we can derive a closed analytic formula for the compression modulus of
the Skyrmions living within such a cuboid. To the best of our knowledge,
this is the first case in which one can derive an analytic formula (Eqs. (%
\ref{Kbox}) and (\ref{Vbox}) below) for the compression modulus in a highly
interacting theory such as the low energy limit of QCD. Indeed, by
expressing the fundamental length as $l=\left( \frac{V}{16\pi ^{3}}\right)
^{1/3}$ we can easily use \eqref{ltox} to relate the volume $V$ with the
variable $x$ on which the energy depends \eqref{encube}. In this manner we
can get an analytical expression for the compression modulus of the cube in
terms of the variable $x$, which is
\begin{equation}
\begin{split}
\mathcal{K}(x)=& -\frac{36}{pq}\Bigg[\left( x^{2}+1\right) \mathrm{K}\left(
-x^{2}\right) ^{3}\left( \pi ^{2}p^{2}q^{2}-x^{2}\mathrm{K}\left(
-x^{2}\right) ^{2}\left( p^{2}+q^{2}\right) \right) ^{2} \\
& +x^{2}\mathrm{K}\left( -x^{2}\right) ^{3}\left( p^{2}+q^{2}\right)
\mathcal{E}\left( -x^{2}\right) ^{2}\left( 5\pi ^{2}p^{2}q^{2}-x^{2}\mathrm{K%
}\left( -x^{2}\right) ^{2}\left( p^{2}+q^{2}\right) \right) \\
& +\pi ^{4}p^{2}q^{2}\mathcal{E}\left( -x^{2}\right) \left( \mathrm{K}\left(
-x^{2}\right) ^{2}\left( x^{2}\left( p^{2}+q^{2}\right)
^{2}-2p^{2}q^{2}\right) -\pi ^{2}p^{2}q^{2}\left( p^{2}+q^{2}\right) \right) %
\Bigg].
\end{split}
\label{Kbox}
\end{equation}%
It can be shown that the parametric plots with respect to the volume which
is
\begin{equation}
V=2\pi ^{3}\frac{\left( \pi ^{2}p^{2}q^{2}-x^{2}\mathrm{K}\left(
-x^{2}\right) ^{2}\left( p^{2}+q^{2}\right) \right) ^{3/2}}{x^{3}\mathrm{K}%
\left( -x^{2}\right) ^{3}}  \label{Vbox}
\end{equation}%
lead to the same behavior for the pressure, the energy and the compression
modulus that has being derived in the previous section. For various values
of $p$ and $q$ the behavior of the before mentioned quantities is described
by the same graphs as given in Fig \ref{Fig5}.

A baryon density ($n=\frac{B}{V}$) of $0.04$ fm$^{-3}$ $\lesssim n\lesssim
0.07$ fm$^{-3}$ is assumed \cite{Caplan} to be appropriate for
characterizing nuclear pasta and in particular lasagna. Within this range
densities we can see that with expressions \eqref{Kbox} and \eqref{Vbox} we
can achieve a compression modulus around $\mathcal{K}\sim 230$MeV (which is
quite reasonable \cite{Adam, Dutra}). For instance in table \ref{tab3} one
can observe various examples of configurations involving baryon densities $n$
and the corresponding baryon numbers $B$, whose compression modulus - as
calculated with the help of \eqref{Kbox} - is $\mathcal{K}\sim 230$MeV. In
all cases presented in the table we have considered $p=q$, thus $B=p^{2}$.

\begin{table}[tbp]
\begin{center}
\begin{tabular}{|c|c|c|c|c|}
\hline
$B$ & 144 & 196 & 225 & 324 \\ \hline
$n$ (fm$^{-3}$) & 0.044 & 0.048 & 0.051 & 0.057 \\ \hline
\end{tabular}%
\end{center}
\caption{Examples of configurations corresponding to a compression modulus $%
\mathcal{K}\sim 230$MeV.}
\label{tab3}
\end{table}

\section{Gauged solitons}

Here we will shortly describe (a slight generalization of) the gauged
solitons constructed in \cite{gaugsk}.

\subsection{Gauged Skyrmions}

As in \cite{gaugsk}, we introduce an electromagnetic potential of the form
\begin{equation}  \label{empot}
A_{\mu }=(b_{1}(r),0,b_{2}(r),b_{3}(r)),
\end{equation}%
to be coupled to the multi-Skyrmionic system under consideration. The
Maxwell equations \eqref{maxwellskyrme1} reduce to
\begin{equation}  \label{Maxbs}
b_{i}^{\prime \prime }=\kappa^2 M_{ij}b_{j}+ \kappa N_{i}
\end{equation}%
with the nonzero components of $M$ and $N$ being
\begin{align*}
M_{11}& =-K\sin ^{2}(H)\left[ l_{1}^{2}\left( 4+\lambda \left( \frac{p^{2}}{%
l_{2}^{2}}+\frac{q^{2}}{l_{3}^{2}}\right) \cos ^{2}(H)\right) +4\lambda
H^{\prime 2}\right] \\
M_{23}& =\frac{K\lambda l_{1}^{2}pq}{4l_{3}^{2}}\sin ^{2}(2H) \\
M_{32}& =\frac{l_{3}^{2}}{l_{2}^{2}}M_{23} \\
M_{22}& =M_{11}+\frac{p}{q}M_{32} \\
M_{22}& =M_{11}+\frac{q}{p}M_{23} \\
N_{2}& =\frac{p}{4}M_{11}+\frac{1}{4}\left( \frac{l_{3}^{2}p^{2}}{l_{2}^{2}q}%
-q\right) M_{23} \\
N_{3}& =-\frac{q}{4}M_{11}-\frac{1}{4}\left( \frac{l_{2}^{2}q^{2}}{l_{3}^{2}p%
}-p\right) M_{32}.
\end{align*}%
A direct computation shows that, using the line element in Eq. (\ref{line3l}%
), the three coupled gauged Skyrme equations (namely, $\mathit{E}^{j}=0$, $%
j=1$, $2$, $3$) in Eq. (\ref{nonlinearsigma1})
\begin{equation*}
D^{\mu }\left( R_{\mu }+\frac{\lambda }{4}\left[ R^{\nu },G_{\mu \nu }\right]
\right) =\mathit{E}^{j}t_{j}=0
\end{equation*}%
reduce to only one Skyrme field equation (since the third Skyrme equation is
identically satisfied while the first and the second are proportional):%
\begin{eqnarray*}
\mathit{E}^{3} &=&0\ , \\
\mathit{E}^{1} &=&I_{1}P\left[ H\right] \ ,\ \mathit{E}^{2}=I_{2}P\left[ H%
\right] \ ,\ \ I_{1}\neq 0\ ,\ I_{2}\neq 0\ ,
\end{eqnarray*}%
where $I_{j}$ are real and non-vanishing. Thus, the Skyrme field equations
reduce to $P\left[ H\right] =0$ namely
\begin{equation}
\begin{split}
& 4\left[ X\sin ^{2}(H)-\lambda \left( l_{2}^{2}q^{2}+l_{3}^{2}p^{2}\right)
-4l_{2}^{2}l_{3}^{2}\right] H^{\prime \prime }+2X\sin (2H)H^{\prime 2}+4\sin
^{2}(H)X^{\prime }H^{\prime } \\
& +\Big[\lambda \kappa \left( l_{3}^{2}pb_{2}+l_{2}^{2}qb_{3}\right) \left( -%
\frac{4l_{1}^{2}p}{l_{2}^{2}}\kappa b_{2}-\frac{4l_{1}^{2}q}{l_{3}^{2}}%
\kappa b_{3}+2l_{1}^{2}\left( \frac{q^{2}}{l_{3}^{2}}-\frac{p^{2}}{l_{2}^{2}}%
\right) \right) -\frac{1}{4}l_{1}^{2}X\left( \frac{p^{2}}{l_{2}^{2}}+\frac{%
q^{2}}{l_{3}^{2}}\right) \\
& +\lambda l_{1}^{2}p^{2}q^{2}\Big]\sin (4H)-\frac{2l_{1}^{2}}{\lambda }%
X\sin (2H)=P\left[ H\right] =0\ ,
\end{split}
\label{profl3}
\end{equation}%
where
\begin{equation}
X(r)=8\lambda \kappa \left( 2l_{2}^{2}l_{3}^{2}\kappa
b_{1}^{2}-l_{3}^{2}b_{2}(2\kappa b_{2}+p)+l_{2}^{2}b_{3}(q-2\kappa
b_{3})\right) .
\end{equation}

Quite remarkably, if we demand that
\begin{equation}
X(r)=\lambda \left( l_{2}^{2}q^{2}+l_{3}^{2}p^{2}\right) ,\quad b_{2}(r)=-%
\frac{l_{2}^{2}q}{l_{3}^{2}p}b_{3}+\frac{1}{\kappa }\left( \frac{%
l_{2}^{2}q^{2}}{4l_{3}^{2}p}-\frac{p}{4}\right) \ ,  \label{Xcon1}
\end{equation}%
then the equation for the profile $H(r)$ can be solved explicitly. More
importantly, the above algebraic conditions in Eq. (\ref{Xcon1}) are
consistent with the Maxwell equations written above. Indeed, if one plugs
the two algebraic conditions in Eq. (\ref{Xcon1}) into the three Maxwell
equations one obtains a single Maxwell equation for $b_{3}(r)$:
\begin{equation}
b_{3}^{\prime \prime }=\frac{\kappa K}{8l_{2}^{2}l_{3}^{2}}(q-4\kappa b_{3})%
\left[ 8\lambda l_{2}^{2}l_{3}^{2}H^{\prime 2}+l_{1}^{2}\left( \lambda \cos
(2H)\left( l_{2}^{2}q^{2}+l_{3}^{2}p^{2}\right) +l_{2}^{2}\left(
8l_{3}^{2}+\lambda q^{2}\right) +\lambda l_{3}^{2}p^{2}\right) \right] \sin
^{2}(H)\ ,  \label{b3XYcon1}
\end{equation}%
while for the profile $H(r)$ we have a decoupled (from $b_{3}$) equation
that reads
\begin{equation}
\left[ \lambda \cos (2H)\left( l_{2}^{2}q^{2}+l_{3}^{2}p^{2}\right)
+l_{2}^{2}\left( 8l_{3}^{2}+\lambda q^{2}\right) +\lambda l_{3}^{2}p^{2}%
\right] H^{\prime \prime }+\left( l_{2}^{2}q^{2}+l_{3}^{2}p^{2}\right)
\left( l_{1}^{2}-\lambda H^{\prime 2}\right) \sin (2H)=0.  \label{profXYcon1}
\end{equation}

Thus, the big technical achievement of the present approach is that the
three coupled gauged Skyrme equations in Eq. (\ref{nonlinearsigma1}) and the
corresponding four Maxwell equations in Eq. (\ref{maxwellskyrme1}) with
exactly the Skyrme ansatz in Eqs. (\ref{pions2.25}) and (\ref{pions2.26})
and the gauge potential in Eq. (\ref{empot}) reduce to Eqs. (\ref{b3XYcon1})
and (\ref{profXYcon1}) when the two algebraic conditions in Eq. (\ref{Xcon1}%
) are satisfied. We want to stress that the aforementioned relations provide
an exact solution and they are not a product of an approximation. As for the
boundary conditions that are needed to be set, we have to keep in mind that
the system is confined to a finite box. Thus, the easiest way to realize
this is by imposing periodic boundary conditions in $\gamma$ and $\phi$ and
Dirichlet in $r$

Interestingly enough, Eq. (\ref{profXYcon1}) can be solved explicitly by
observing that it has the following first integral%
\begin{equation}
Y\left( H\right) \frac{H^{\prime 2}}{2}+V(H)=E_{0}\ ,  \label{firstint}
\end{equation}%
with
\begin{eqnarray}
Y\left( H\right) &=&2\lambda \left( l_{2}^{2}q^{2}+l_{3}^{2}p^{2}\right)
\cos ^{2}(H)+8l_{2}^{2}l_{3}^{2},  \label{firstint2} \\
V\left( H\right) &=&-\frac{1}{2}l_{1}^{2}\left(
l_{2}^{2}q^{2}+l_{3}^{2}p^{2}\right) \cos (2H)  \label{firstint3}
\end{eqnarray}%
and where $E_{0}$ is an integration constant to be determined by requiring
that the boundary conditions to have non-vanishing topological charge are
satisfied. Thus, Eq. (\ref{profXYcon1}) can be reduced to a quadrature
(which defines a generalized elliptic integral). Eq. (\ref{b3XYcon1}) for $%
b_{3}$ is linear (since $H(r)$ can be found explicitly), however its
integration is not a trivial task. In any case, integration of (\ref%
{b3XYcon1}) that results in an expression for $b_{3}$ makes trivial the
determination of the other two components of $A_{\mu }$ since both $b_{1}$
and $b_{2}$ are given algebraically in terms of $b_{3}$ through conditions %
\eqref{Xcon1}. Nevertheless, even without the explicit expressions, it is
still possible to analyze the generic features of the transport properties
electrons passing through the above gauged Skyrmions.

\subsection{Gauged time-crystals}

In order to have a time periodic solution with a non vanishing topological
charge, that can be characterized as a time-crystal (for the introduction to
the notion of time crystals see \cite{timec1,timec2,timec3,timecr}) we start
by considering the line element
\begin{equation}
ds^2 = - d\gamma^2 +l_1 dr^2 +l_2 dz^2 +l_3 d\phi^2,
\end{equation}
where $\gamma$ in the new ansatz
\begin{equation}
G=\frac{q\phi - \omega \gamma}{2}, \; A=- \frac{q\phi + \omega \gamma}{2}
\end{equation}
is the time variable, making the ensuing solution a time periodic
configuration. The constant $\omega$ is the frequency of the time-crystal
characterizing the periodicity of the system. Again we consider a finite
box, where this time we take
\begin{equation}
0 \leq r \leq 2\pi, \quad 0 \leq z \leq 4\pi, \quad 0 \leq \phi \leq 2\pi .
\end{equation}
We adopt a similar form for the electromagnetic potential as the one given
in \eqref{empot}. However, we have to note now that the index of the
coordinates is changed into $x^\mu = (\gamma, r, z, \phi)$. Thus, the vector
potential is
\begin{equation}
A_\mu = (b_2(r),0,b_1(r),b_3(r)),
\end{equation}
making $b_2(r)$ the electrostatic potential instead of $b_1(r)$ that we had
in the Skyrmion case. The Maxwell equations \eqref{maxwellskyrme1} retain
same form as \eqref{Maxbs} with
\begin{align*}
M_{11} & = -\frac{K}{2 l_3^2} \sin ^2(H) \left[8 \lambda l_3^2 H^{\prime
2}+l_1^2 \left(2 \lambda \cos ^2(H) \left(q^2-l_3^2 \omega ^2\right)+8
l_3^2\right)\right] \\
M_{23} & = \frac{K \lambda l_1^2 q \omega}{4 l_3^2} \sin ^2(2 H) \\
M_{32} & = - l_3^2 M_{23} \\
M_{22} & = M_{11} + \frac{\omega}{q} M_{3,2} \\
M_{33} & = M_{11} + \frac{q}{\omega} M_{2,3} \\
N_2 & = \frac{\omega}{4} M_{11} - \frac{1}{4}\left(\frac{l_3^2 \omega ^2}{q}
+q\right) \\
N_3 & = -\frac{q}{4}M_{11} + \frac{1}{4} \left(\frac{q^2}{l_3^2 \omega } +
\omega \right),
\end{align*}
while the rest of the components of $M$ and $N$ are zero.

As also happened in the Skyrme case, again here, the field equations reduce
to a single ordinary differential equation for the profile function $H(r)$.
In this case the relative equation reads
\begin{equation}
\begin{split}
& 4\left( X\sin ^{2}(H)+l_{2}^{2}\left( l_{3}^{2}\left( \lambda \omega
^{2}-4\right) -\lambda q^{2}\right) \right) H^{\prime \prime }+2X\sin
(2H)(H^{\prime })^{2}+4\sin ^{2}(H)X^{\prime }H^{\prime } \\
& +\frac{l_{1}^{2}}{4l_{3}^{2}}\left[ 4\lambda l_{2}^{2}\left( 2\kappa
qb_{3}-l_{3}^{2}\omega (2\kappa b_{2}+\omega )\right) \left( 2\kappa
l_{3}^{2}\omega b_{2}+q(q-2\kappa b_{3})\right) -X\left(
q^{2}-l_{3}^{2}\omega ^{2}\right) \right] \sin (4H) \\
& -\frac{\left( 2l_{1}^{2}\right) }{\lambda }X\sin (2H)=0,
\end{split}
\label{profeqTC}
\end{equation}%
where
\begin{equation}
X(r)=-8\kappa \lambda \left[ 2\kappa l_{3}^{2}b_{1}^{2}-l_{2}^{2}\left(
l_{3}^{2}b_{2}(2\kappa b_{2}+\omega )+b_{3}(q-2\kappa b_{3})\right) \right] .
\end{equation}%
Once more, profile equation \eqref{profeqTC} can be reduced to an integrable
one that is decoupled from the Maxwell field. Let us assume the following
conditions for the components $b_{1}$ and $b_{3}$ of the electromagnetic
potential $A_{\mu }$:
\begin{equation}
X(r)=\lambda l_{2}^{2}\left( q^{2}-l_{3}^{2}\omega ^{2}\right) ,\quad
b_{3}(r)=\frac{l_{3}^{2}\omega }{q}b_{2}(r)+\frac{l_{3}^{2}\omega ^{2}}{%
4\kappa q}+\frac{q}{4\kappa }.
\end{equation}%
Then, the remaining Maxwell equation that needs to be satisfied for $b_{2}$
is
\begin{equation}
b_{2}^{\prime \prime }=-\frac{\kappa K}{8l_{3}^{2}}(4\kappa b_{2}+\omega )%
\left[ 8l_{3}^{2}\left( \lambda (H^{\prime })^{2}+l_{1}^{2}\right) +2\lambda
l_{1}^{2}\cos ^{2}(H)\left( q^{2}-l_{3}^{2}\omega ^{2}\right) \right] \sin
^{2}(H)  \label{matc}
\end{equation}%
and the profile equation is reduced to
\begin{equation}
\left( 2\lambda \cos ^{2}(H)\left( q^{2}-l_{3}^{2}\omega ^{2}\right)
+8l_{3}^{2}\right) H^{\prime \prime }+\sin (2H)\left( q^{2}-l_{3}^{2}\omega
^{2}\right) \left( l_{1}^{2}-\lambda H^{\prime 2}\right) =0.
\end{equation}%
Obviously it exhibits a first integral of the form \eqref{firstint} where
now
\begin{align*}
Y(H)& =2\lambda \cos ^{2}(H)\left( q^{2}-l_{3}^{2}\omega ^{2}\right)
+8l_{3}^{2} \\
V(H)& =\frac{l_{1}^{2}}{2}\left( l_{3}^{2}\omega ^{2}-q^{2}\right) \cos (2H).
\end{align*}%
We can notice the similarities with the expressions derived for the Skyrmion
in the previous case. In \cite{gaugsk} there has been presented an extensive
discussion on the \textquotedblleft extended duality" that exists between
two such systems.

\subsection{Topological Current for the gauged Skyrmion}

The topological current \cite{Witten} of the gauged Skyrme model can be
divided into two terms
\begin{equation}
J_{\mu }^{B}=J_{\mu }^{Sk}+J_{\mu }^{B-em}
\end{equation}%
with the first term $J_{\mu }^{Sk}$ being the usual Baryonic current, while
second term is the correction to the latter, owed to the coupling with the
electromagnetic field. For the first term we have
\begin{equation}
J_{\mu }^{Sk}=\frac{1}{24\pi ^{2}}E_{\mu \alpha \beta \nu }Tr\left(
R^{\alpha }R^{\beta }R^{\nu }\right) ,  \label{Dirac}
\end{equation}%
which in our case has a single nonzero component
\begin{equation}
J_{0}^{Sk}=-\frac{pq}{8\pi ^{2}l_{1}l_{2}l_{3}}H^{\prime }\sin (2H)=-2\pi
\widehat{n}_{B}H^{\prime }\sin (2H)\ ,\ V=16\pi ^{3}l_{1}l_{2}l_{3},
\label{Bden1}
\end{equation}%
where $V=16\pi ^{3}l_{1}l_{2}l_{3}$ is the volume of the box and $\widehat{n}%
_{B}$\ is the Baryon density ($\widehat{n}_{B}=pq/V$) of the system. Note
that in \eqref{Dirac} we make use of the Levi-Civita tensor $E_{\mu \alpha
\beta \nu }=\sqrt{-g}\,\epsilon _{\mu \alpha \beta \nu }$ instead of the
Levi-Civita symbol $\epsilon _{\mu \alpha \beta \nu }$ so that $J_{\mu
}^{Sk} $ transforms covariantly and the topological charge results in a pure
number. If for instance we apply the boundary conditions $H(0)=0$, $H(2\pi )=%
\frac{\pi }{2}$ we obtain
\begin{equation}
B=\int_{\Sigma }\!\!\sqrt{-g}J_{Sk}^{0}drd\gamma d\phi =pq.
\end{equation}%
The correction $J_{\mu }^{B-em}$ to the baryonic current, due to the
electromagnetic field, is
\begin{equation}
J_{\mu }^{em}=-\frac{\kappa }{8\pi ^{2}}E_{\mu \alpha \beta \nu }\nabla
^{\alpha }\left[ A^{\beta }Tr\left( t_{3}(U^{-1}\nabla ^{\nu }U-\nabla ^{\nu
}UU^{-1})\right) \right]  \label{curBem}
\end{equation}%
and the total gauged Baryonic current reads
\begin{equation}  \label{fullcurB}
\begin{split}
J_{\mu }^{B}=\Big\{-\frac{pq\pi }{V}\partial _{r}\left( \cos (2H)\right) +%
\frac{4\pi \kappa }{V}\partial _{r}\left( \cos ^{2}(H)(qb_{2}-pb_{3})\right)
,0,& \\
-\frac{4\pi q\kappa }{V}\partial _{r}\left( b_{1}\cos ^{2}(H)\right) ,\frac{%
4\pi p\kappa }{V}\partial _{r}\left( \cos ^{2}(H)\right) & \Big\},
\end{split}%
\end{equation}%
From what we see, the total baryon number when the Skyrmion is coupled to
the electromagnetic field depends also on the boundary conditions that one
may impose on the latter ($b_{2}$ and $b_{3}$ in particular).

\subsection{Baryonic current for the Time-Crystal}

The topological current of the time-crystal can be calculated with the use
of the same relations \eqref{Dirac} and \eqref{curBem}. Here we just give
the result for the full current of the Gauged Time Crystal (GTC) which is
\begin{equation}  \label{JGTC}
\begin{split}
J_\mu^{GTC} = \Big\{ &-\frac{4\pi q \kappa }{V} \partial_r \left(b_1
\cos^2(H)\right), 0, \\
& -\frac{l_2^2 q \pi \omega }{ V} \partial_r (\cos(2H)) + \frac{4 \pi \kappa
}{V} \partial_r\left[\cos ^2(H) (q b_2-\omega b_3)\right], \frac{4\pi \kappa
\omega }{V} \partial_r \left(b_1 \cos^2(H) \right) \Big\} .
\end{split}%
\end{equation}
In the absence of the coupling with the electromagnetic field, $\kappa=0$,
we can see that the expression for the non-zero topological current of the
time-crystal is simplified to
\begin{equation}
J_\mu^{TC} = \Big\{ 0,0, -\frac{\pi l_2^2 q \omega }{V} \partial_r
(\cos(2H)),0\Big\} .
\end{equation}

\section{On the conductivity of gauged solitons}

At semi-classical level, the transport properties of electrons travelling
through the above gauged Skyrmions can be determined by analyzing the
corresponding Dirac equation. Obviously, the electrons interact directly
both with the gauge field and with the Baryons. The fermion couples to $%
A_{\mu }$, as QED dictates. However, there are further effects due to the
coupling with the baryonic current. Here, we follow a very simple toy model
interaction just to make a qualitative description of such effects. At this
level of approximation in which the electrons perceive the gauged Skyrmions
as a classical background, both interactions can be described as
\textquotedblleft current-current" interactions in the Dirac Hamiltonian.
The interaction of the electronic Dirac field $\Psi $ with the gauge
potential $A_{\mu }$ corresponds to the following interaction Hamiltonian%
\begin{eqnarray}
H_{int}^{U(1)} &=&\kappa J_{\mu }^{e}A^{\mu }\ ,  \label{u1corr} \\
J_{\mu }^{e} &=&\overline{\Psi }\gamma _{\mu }\Psi \ ,  \notag \\
\overline{\Psi } &=&\Psi ^{\dag }\gamma ^{0},  \notag
\end{eqnarray}%
where $\kappa $\ is the Maxwell coupling%
\begin{equation}
\kappa \approx \left( \frac{1}{137}\right) ^{\frac{1}{2}},
\label{u1coupling}
\end{equation}%
$\gamma _{\mu }$ are the Dirac gamma-matrices (the conventions are collected
in the appendix \ref{appA}), $\Psi ^{\dag }$ is the conjugate transpose of $%
\Psi $ and $\overline{\Psi }$ the adjoint spinor. On the other hand, a
simple way to describe the interactions of the electronic Dirac field with
the baryonic current $J_{B}^{\mu }$ is with the following Hamiltonian%
\begin{equation}
H_{int}^{B}=g_{eff}J_{\mu }^{e}J_{B}^{\mu }\ ,  \label{barcorr}
\end{equation}%
where $g_{eff}$\ is the effective coupling constant of the electron-Baryon
interaction. At the present level of approximation (in which the energy
scale is not high enough to disclose the parton structure of the Baryon) a
reasonable assumption is:
\begin{equation*}
g_{eff}\approx G_{F},
\end{equation*}%
where $G_{F}$ is the Fermi constant.

In order to evaluate the relative strength of the two contributions to the
conductivity (a brief analysis is given in Appendix \ref{appB}), one arising
from the term owed to the coupling with the $U(1)$ field (the $\kappa A_{\mu
}$\ in Eq. (\ref{Diraceq}), see section \ref{appB1} of Appendix \ref{appB})
and the other arising from the term produced from the baryon current (the $%
G_F J_{\mu }^{B}$\ in Eq. (\ref{Diraceq})) one needs to evaluate the
relative strength of the $U(1)$ coupling with respect to the interactions
with the Skyrmionic current. There are two competing factors in the
interactions with the Skyrmionic current. The first factor is the
electro-weak coupling constant (which is obviously weaker than the $U(1)$
coupling). The second factor is related with the Skyrmions profile $H$ and
can be evaluated explicitly thanks to the present analytic solutions.
Assuming that both $\sin (2H)$ and $H^{\prime }$ are of order 1 (since both
quantities are adimensional and the solitonic solutions we are considering
are smooth and regular) one can see that the effective adimensional coupling
$\widehat{g}$ measuring the strength of the contributions to the
conductivity due to the interactions of the electrons with the Skyrmionic
current is:
\begin{equation}
\widehat{g}= l_1 G_{F}\widehat{n}_{B}\ .  \label{efcoupl}
\end{equation}%
Given that $G_F \sim 1.166$ GeV$^{-2}$ or $G_F \sim 4.564$ fm$^2$ in natural
units we can see that the contribution of the interaction with $J_{\mu }^{B}$
remains small in comparison to the coupling with $A_{\mu }$ - at least for
baryon densities $\widehat{n}_{B}$ and lengths $l_1$ of the box that can be
characterized as natural. The ``Baryonic" correction $\delta \Psi $ to the
wave function in Eq. (\ref{perturbedpsi}) depends on the effective coupling $%
\widehat{g}$\ defined in Eq. (\ref{efcoupl}) and on the Fourier transform of
quantities related with the background Skyrmion.

For completeness, in sections \ref{appB2} and \ref{appB3} of Appendix \ref%
{appB} we have included the Dirac equations for the electrons propagating in
the gauged solitons background described above. Although these Dirac
equations cannot be solved analytically (due to the fact that Eqs. (\ref%
{b3XYcon1}) and (\ref{matc}) are not integrable in general), they can be
useful starting points for numerical analysis of transport properties of the
present gauged solitons.

\section{Conclusions and perspectives}

\label{conclusions}

In the present paper we have studied (gauged) Skyrmionic configurations in a
finite box. We provided the reduced field equations under the adopted ansatz
and distinguished the conditions over the potential functions $A_\mu$ for
which the aforementioned equations can be characterized as integrable.
Additionally, we have presented analytic expressions for the energy and
studied its general behaviour in relation to the baryon number and the
possible sizes of the box under consideration. We also managed to
demonstrate and analyze the cases where the more energetically convenient
configurations emerge in relations to these variables.

What is more, we have derived an explicit analytic expression for the
compression modulus corresponding to Skyrmions living within a finite volume
in flat space-times. This is the first case in which one can derive an
analytic formula (Eqs. (\ref{Kbox}) and (\ref{Vbox}) in the previous
section) for such an important quantity in a highly interacting theory such
as the low energy limit of QCD. This expression produces a
reasonable value with a correct order of magnitude. The gauged version of
these solitons living within a finite volume can be also considered. Using
these gauged solitons, it is possible to analyze the contributions to the
electrons conductivity associated to the interactions with this Baryonic
environment (which represents a slab of baryons which can be very large in
two of the three spatial directions). To the best of authors knowledge, the
present is the first concrete setting in which it is possible to perform
analytic computations of these relevant quantities in the original version
of the Skyrme model (and its gauged version).

\subsection*{Acknowledgements}

The authors would like to thank A. Zerwekh for useful discussions. This work
has been funded by the Fondecyt grants 1160137, 1161150 and 3160121. The
Centro de Estudios Cient\'{\i}ficos (CECs) is funded by the Chilean
Government through the Centers of Excellence Base Financing Program of
Conicyt.

\appendix

\section{Conventions}

\label{appA}

Throughout the paper we use the metric signature $(-,+,+,+)$. The ordering
of the space-time coordinates is $x^{\mu }=(z,r,\gamma ,\phi )$ for the
Skyrmion and $x^{\mu }=(\gamma ,r,z,\phi )$ for the time-crystal.

The four Dirac matrices are
\begin{align*}
\gamma ^{0}& =\left(
\begin{array}{cccc}
1 & 0 & 0 & 0 \\
0 & 1 & 0 & 0 \\
0 & 0 & -1 & 0 \\
0 & 0 & 0 & -1%
\end{array}%
\right) ,\quad \gamma ^{1}=\left(
\begin{array}{cccc}
0 & 0 & 0 & 1 \\
0 & 0 & 1 & 0 \\
0 & -1 & 0 & 0 \\
-1 & 0 & 0 & 0%
\end{array}%
\right) \\
\gamma ^{2}& =\left(
\begin{array}{cccc}
0 & 0 & 0 & -\mathbbmtt{i} \\
0 & 0 & \mathbbmtt{i} & 0 \\
0 & \mathbbmtt{i} & 0 & 0 \\
-\mathbbmtt{i} & 0 & 0 & 0%
\end{array}%
\right) ,\quad \gamma ^{3}=\left(
\begin{array}{cccc}
0 & 0 & 1 & 0 \\
0 & 0 & 0 & -1 \\
-1 & 0 & 0 & 0 \\
0 & 1 & 0 & 0%
\end{array}%
\right) .
\end{align*}

\section{Dirac equation}

\label{appB}

Here we include, for completeness, the Dirac equation for an electron
propagating in the two gauged solitons described in the main text. Although,
in these cases, the Dirac equation cannot be solved analytically, it shows
clearly that the present framework provides with a concrete setting to
attack computations which, at a first glance, could appear very difficult
(like the conductivities associated to gauged solitons at finite densities).

\subsection{Qualitative Analysis}

\label{appB1}

The Dirac equation which describes the propagation of the electron through
the above gauged Skyrmion is
\begin{equation}
\left[ \gamma ^{\mu }\left( \mathbbmtt{i}\nabla _{\mu }-\kappa A_{\mu
}-G_{F}J_{\mu }^{B}\right) +m\right] \Psi (z,r,\gamma ,\phi )=0\ ,
\label{Diraceq}
\end{equation}%
where $m$ is the electron mass and $J_{\mu }^{B}$ is given by %
\eqref{fullcurB}. It is convenient to write the above Dirac equation as
follows\footnotetext{%
On the other hand, the gauge potential $A_{\mu }$ and the Baryonic current $%
J_{\mu }^{B}$ are the ones corresponding to the gauged Skyrmion and gauged
time-crystal described in the previous section.}:%
\begin{eqnarray}
\left( H_{0}+H_{int}\right) \Psi &=&0\ ,  \label{int0} \\
H_{0} &=&\left[ \mathbbmtt{i}\gamma ^{\mu }\nabla _{\mu }+m\right] \ ,
\label{int1} \\
H_{int} &=&\left[ \gamma ^{\mu }\left( -\kappa A_{\mu }-G_{F}J_{\mu
}^{B}\right) \right] \footnotemark .  \label{int3}
\end{eqnarray}%
We will work to first order in perturbation theory and we will consider $%
H_{int}$ as a small perturbation. The main goal of our analysis is to take
the first order corrections to the conductivity and make a comparison
between the part that is owed to the interactions with the solitons and the
usual contributions arising from electromagnetic sources other than the
soliton itself.

The last \textit{ingredient} we need is the Kubo formula for the
conductivity associated to electrons moving in a medium (for a detailed
review see chapter 4 of \cite{Dressel}). Following the usual steps one
arrives at the following expression for the conductivity $\sigma _{\mu \nu
}\left( \overrightarrow{q},\Omega \right) $ (where $\overrightarrow{q}$ and $%
\Omega$ the wave vector and frequency respectively of the incident
electromagnetic wave):
\begin{equation*}
\sigma _{\mu \nu }\left( \overrightarrow{q},\Omega \right) =\sum_{s}\frac{1}{%
\hbar \Omega }\int dt\left\langle s\right\vert J_{0\mu }^{e}\left(
\overrightarrow{q},0\right) J_{0\nu }^{\ast e}\left( \overrightarrow{q}%
,\Omega \right) \left\vert s\right\rangle \exp \left[ -i\Omega t\right]
\end{equation*}%
where $\left\vert s\right\rangle $\ and $J_{0\mu }^{e}$ are the eigenstate
of the free Dirac Hamiltonian and the corresponding current in the box where
the gauged solitons live.

Due to the interaction Hamiltonian $H_{int}$\ defined\footnote{%
The gauge potential $A_{\mu }$ and the Baryon current $J_{\mu }^{B}$ in the
interaction Hamiltonian are the ones corresponding to the gauged Skyrmion
and to the gauged time-crystal defined in the previous section.} in Eqs. (%
\ref{int0}), (\ref{int1}) and (\ref{int3}), the electron currents $J_{\mu
}^{e}=\overline{\Psi }\gamma _{\mu }\Psi $ changes%
\begin{equation*}
J_{0\mu }^{e}\rightarrow J_{0\mu }^{e}+\left( \delta \overline{\Psi }\right)
\gamma _{\mu }\Psi +\overline{\Psi }\gamma _{\mu }\left( \delta \Psi \right)
=J_{0\mu }^{e}+\delta J_{\mu }^{e}\ ,
\end{equation*}%
where $\delta \Psi $\ can be computed using first order perturbation theory.
In particular, if $\Psi _{0}$ is a solution of the un-perturbed equation%
\begin{equation*}
H_{0}\Psi _{0}=E\Psi _{0}\ ,
\end{equation*}%
then the eigenstate $\Psi $ of the interacting case can be written as%
\begin{equation*}
\Psi =\Psi _{0}-H_{0}^{-1}\left( H_{int}\Psi _{0}\right) ,
\end{equation*}%
where $H_{0}^{-1}$ is the inverse Dirac operator defined as the Green
function $H_{0}^{-1}=G(x-x^{\prime })$ satisfying
\begin{equation}
H_{0}G_{0}(x-x^{\prime })=\delta (x-x^{\prime }).
\end{equation}%
We now from the free particle case that the Green function in space-time
variables is expressed as
\begin{equation}
H_{0}^{-1}=G_{0}(x-x^{\prime })=\int \!\!\frac{d^{4}k}{(2\pi )^{4}}e^{-%
\mathbbmtt{i}k_{\mu }(x^{\mu }-x^{\prime \mu })}\frac{m-\gamma ^{\mu }k_{\mu
}}{k_{\mu }k^{\mu }+m^{2}}
\end{equation}%
(of course in our case, for the finite box, the integral is to be
substituted by series). Consequently, we have a perturbation of the form
\begin{equation}
\delta \Psi =\int \!\!\frac{d^{4}k}{(2\pi )^{4}}\int d^{4}x^{\prime -%
\mathbbmtt{i}k_{\mu }(x^{\mu }-x^{\prime \mu })}\frac{m-\gamma ^{\mu }k_{\mu
}}{k_{\mu }k^{\mu }+m^{2}}\left( \kappa A_{\mu }+G_{F}J_{\mu }^{B}\right)
\Psi _{0}(x^{\prime })  \label{perturbedpsi}
\end{equation}%
owed to two contributions; the Maxwell field $A_{\mu }$ and the baryon
current $J_{\mu }^{B}$.

As for the free particle solution $\Psi _{0}$, it is easy to see that
\begin{equation}
\Psi _{0}(x)=%
\begin{pmatrix}
\psi _{1} \\
\psi _{2} \\
\psi _{3} \\
\psi _{4}%
\end{pmatrix}%
e^{-\mathbbmtt{i}k_{\mu }x^{\mu }}
\end{equation}%
with
\begin{equation}
\psi _{1}=\frac{k_{3}\psi _{3}+(k_{1}-\mathbbmtt{i}k_{2})\psi _{4}}{k_{0}+m}%
,\quad \psi _{2}=\frac{(k_{1}+\mathbbmtt{i}k_{2})\psi _{3}-k_{3}\psi _{4}}{%
k_{0}+m},\quad k_{0}^{2}=\vec{k}^{2}+m
\end{equation}%
satisfies $H_{0}\Psi _{0}=0$.

Consequently,
\begin{eqnarray*}
\sigma _{\mu \nu } &\rightarrow &\sigma _{\mu \nu }+\delta \sigma _{\mu \nu
}\ , \\
\delta \sigma _{\mu \nu } &=&\sum_{s}\frac{1}{\hbar \Omega }\int
dt\left\langle s\right\vert \left[ \delta J_{0\mu }^{e}\left(
\overrightarrow{q},0\right) J_{0\nu }^{\ast e}\left( \overrightarrow{q}%
,\Omega \right) +J_{0\mu }^{e}\left( \overrightarrow{q},0\right) \delta
J_{0\nu }^{\ast e}\left( \overrightarrow{q},\Omega \right) \right]
\left\vert s\right\rangle \exp \left[ -i\Omega t\right] \ .
\end{eqnarray*}

\subsection{Dirac equation for the gauged Skyrmion}

\label{appB2}

The symmetries of the problem allow to search for a separated solution of
the form
\begin{equation}  \label{Danz1}
\Psi (z,r,\gamma ,\phi )=e^{-\mathbbmtt{i}(\omega _{1}z-k_{2}\gamma
-k_{3}\phi )}\{\psi _{1}(r),\psi _{2}(r),\psi _{3}(r),\psi _{4}(r)\}.
\end{equation}%
By introducing \eqref{Danz1} into the Dirac equation \eqref{Diraceq} we
obtain the following set of equations for the components of $\Psi$:
\begin{subequations}
\begin{align}
\psi _{1}^{\prime }=& \left( k_{2}+\kappa b_{2}-\kappa g\frac{q}{4\pi ^{2}V}%
\phi _{1}^{\prime }\right) \psi _{1}+\mathbbmtt{i}\left( k_{3}+\kappa
b_{3}+\kappa g\frac{p}{4\pi ^{2}V}\phi _{1}^{\prime }\right) \psi _{2}
\notag \\
& +\mathbbmtt{i}\left( \omega _{1}-m-\kappa b_{1}-\frac{g}{16\pi ^{2}V}%
\left( 4\kappa \phi _{2}^{\prime }+\phi _{3}^{\prime }\right) \right) \psi
_{4} \\
\psi _{2}^{\prime }=& -\mathbbmtt{i}\left( \kappa b_{3}+k_{3}+\kappa g\frac{p%
}{4\pi ^{2}V}\phi _{1}^{\prime }\right) \psi _{1}-\left( \kappa
b_{2}+k_{2}-\kappa g\frac{q}{4\pi ^{2}V}\phi _{1}^{\prime }\right) \psi _{2}
\notag \\
& +\mathbbmtt{i}\left( \omega _{1}-m-\kappa b_{1}-\frac{g}{16\pi ^{2}V}%
\left( 4\kappa \phi _{2}^{\prime }+\phi _{3}^{\prime }\right) \right) \psi
_{3} \\
\psi _{3}^{\prime }=& \mathbbmtt{i}\left( \omega _{1}+m-\kappa b_{1}-\frac{g%
}{16\pi ^{2}V}\left( 4\kappa \phi _{2}^{\prime }+\phi _{3}^{\prime }\right)
\right) \psi _{2}  \notag \\
& +\left( k_{2}+\kappa b_{2}-\kappa g\frac{q}{4\pi ^{2}V}\phi _{1}^{\prime
}\right) \psi _{3}+\mathbbmtt{i}\left( k_{3}+\sigma b_{3}+\kappa g\frac{p}{%
4\pi ^{2}V}\phi _{1}^{\prime }\right) \psi _{4} \\
\psi _{4}^{\prime }=& \mathbbmtt{i}\left( \omega _{1}+m-\kappa b_{1}-\frac{g%
}{16\pi ^{2}V}\left( 4\kappa \phi _{2}^{\prime }+\phi _{3}^{\prime }\right)
\right) \psi _{1}  \notag \\
& -\mathbbmtt{i}\left( \kappa b_{3}+k_{3}+\kappa g\frac{p}{4\pi ^{2}V}\phi
_{1}^{\prime }\right) \psi _{3}-\left( \kappa b_{2}+k_{2}-\kappa g\frac{q}{%
4\pi ^{2}V}\phi _{1}^{\prime }\right) \psi _{4},
\end{align}%
where
\end{subequations}
\begin{align}
\phi _{1}(r)& =b_{1}(r)\cos ^{2}(H(r))\ , \\
\phi _{2}(r)& =\cos ^{2}(H(r))(qb_{2}(r)-pb_{3}(r))\ , \\
\phi _{3}(r)& =pq\cos (2H(r))\ .
\end{align}

\subsection{Dirac equation for the gauged time-crystal}

\label{appB3}

By using the expression for $J_{\mu }^{GTC}$ as given by \eqref{JGTC} inside %
\eqref{Diraceq}, instead of $J_{\mu }^{B}$ that we had for the Skyrmion, and
by considering a separable solution of the form
\begin{equation}
\Psi (z,r,\gamma ,\phi )=e^{-\mathbbmtt{i}(\omega _{1}\gamma
-k_{2}z-k_{3}\phi )}\{\psi _{1}(r),\psi _{2}(r),\psi _{3}(r),\psi _{4}(r)\},
\end{equation}%
we obtain a system of equations given by
\begin{subequations}
\begin{align}
\psi _{1}^{\prime }=& \left( \kappa b_{2}-\omega _{1}-\frac{gq\kappa }{4\pi
^{2}V}\phi _{1}^{\prime }\right) \psi _{1}+\mathbbmtt{i}\left( \kappa
b_{3}+k_{3}+\frac{g\kappa \omega }{4\pi ^{2}V}\phi _{1}^{\prime }\right)
\psi _{2}  \notag \\
& -\mathbbmtt{i}\left( \kappa b_{1}+k_{2}+m+\frac{g}{16\pi ^{2}V}\left(
4\kappa \phi _{2}^{\prime }-l_{2}^{2}\phi _{3}^{\prime }\right) \right) \psi
_{4} \\
\psi _{2}^{\prime }=& -\mathbbmtt{i}\left( \kappa b_{2}+k_{3}+\frac{g\kappa
\omega }{4\pi ^{2}V}\phi _{1}^{\prime }\right) \psi _{1}+\left( \omega
_{1}-\kappa b_{2}+\frac{gq\kappa }{4\pi ^{2}V}\phi _{1}^{\prime }\right)
\psi _{2}  \notag \\
& -\mathbbmtt{i}\left( \kappa b_{1}+k_{2}+m+\frac{g}{16\pi ^{2}V}\left(
4\kappa \phi _{2}^{\prime }-l_{2}^{2}\phi _{3}^{\prime }\right) \right) \psi
_{3} \\
\psi _{3}^{\prime }=& \mathbbmtt{i}\left( m-\kappa b_{1}-k_{2}-\frac{g}{%
16\pi ^{2}V}\left( 4\kappa \phi _{2}^{\prime }-l_{2}^{2}\psi _{3}^{\prime
}\right) \right) \psi _{2}  \notag \\
& \left( \kappa b_{2}-\frac{gq\kappa }{4\pi ^{2}V}\phi _{1}^{\prime }-\omega
_{1}\right) \psi _{3}+\mathbbmtt{i}\left( \kappa b_{3}+k_{3}+\frac{g\kappa
\omega }{4\pi ^{2}V}\phi _{1}^{\prime }\right) \psi _{4} \\
\psi _{4}^{\prime }=& \mathbbmtt{i}\left( m-\kappa b_{1}-k_{2}-\frac{g}{%
16\pi ^{2}V}\left( 4\kappa \phi _{2}^{\prime }-l_{2}^{2}\phi _{3}^{\prime
}\right) \right) \psi _{1}  \notag \\
& -\mathbbmtt{i}\left( \kappa b_{3}+k_{3}+\frac{g\kappa \omega }{4\pi ^{2}V}%
\phi _{1}^{\prime }\right) \psi _{3}+\left( \omega _{1}-\kappa b_{2}+\frac{%
gq\kappa }{4\pi ^{2}V}\phi _{1}^{\prime }\right) \psi _{4}.
\end{align}%
The functions $\phi _{1}(r)$, $\phi _{2}(r)$ and $\phi _{3}(r)$ are the same
as before, only now we have $\omega $ appearing in them in place of $p$,
i.e.
\end{subequations}
\begin{align}
\phi _{1}(r)& =b_{1}(r)\cos ^{2}(H(r)) \\
\phi _{2}(r)& =\cos ^{2}(H(r))(qb_{2}(r)-\omega b_{3}(r)) \\
\phi _{3}(r)& =q\omega \cos (2H(r)).
\end{align}

\end{document}